\documentclass[reprint,aps,pra,eqsecnum]{revtex4-1}
\usepackage{amsmath}
\usepackage{amssymb}
\usepackage{amsfonts}\usepackage{color}
\usepackage{graphicx}
\usepackage{epstopdf}

\DeclareMathOperator{\sinc}{sinc}
\DeclareMathOperator{\Tr}{Tr}

\begin{document}

\title{Spontaneous parametric downconversion in waveguides: What's loss
got to do with it?}

\author{L. G. Helt}

\affiliation{Centre for Ultrahigh bandwidth Devices for Optical Systems (CUDOS),
MQ Photonics Research Centre, Department of Physics and Astronomy,
Macquarie University, NSW 2109, Australia}

\author{J. E. Sipe}

\affiliation{Department of Physics and Institute for Optical Sciences, University
of Toronto, 60 St. George St., Toronto, ON M5S 1A7, Canada}

\author{M. J. Steel}

\affiliation{Centre for Ultrahigh bandwidth Devices for Optical Systems (CUDOS),
MQ Photonics Research Centre, Department of Physics and Astronomy,
Macquarie University, NSW 2109, Australia}

\begin{abstract}
We derive frequency correlation and exit probability expressions for photons generated via spontaneous parametric downconversion (SPDC) in nonlinear waveguides that exhibit linear scattering loss. Such loss is included within a general Hamiltonian formalism by connecting waveguide modes to reservoir modes with a phenomenological coupling Hamiltonian, the parameters of which are later related to the usual loss coefficients. In the limit of a low probability of SPDC pair production, the presence of loss requires that we write the usual lossless generated pair state as a reduced density operator, and we find that this density operator is naturally composed of two photon, one photon, and zero photon contributions.  The biphoton probability density, or joint spectral intensity (JSI), associated  with the two-photon contribution is determined not only by a phase matching term, but also by a loss matching term.  The relative sizes of the loss coefficients within this term lead to three qualitatively different regimes of SPDC JSIs. If  either the pump or generated photon loss is much higher than the other, the side lobes of the phase matching squared sinc function are washed out.  On the other hand, if pump and generated photon loss are appropriately balanced, the lossy JSI is identical to the lossless JSI.  Finally, if the generated photon loss is frequency dependent, the shape of the JSI can be altered more severely, potentially leading to generated photons that are less frequency correlated though also produced less efficiently when compared to photons generated in low-loss waveguides. 
\end{abstract}


\maketitle

\section{Introduction}

Waveguides are fundamental integrated optical components. Their performance is key to realizing increased miniaturization, stability, and scalability of both classical and quantum optical devices. In particular, as nonlinear quantum optics experiments continue to move from bulk crystal optics to chip-scale optics, waveguide losses will have a direct effect on nonlinear optical photon generation and manipulation. Indeed, loss mechanisms in waveguides have been well-investigated both theoretically and experimentally~\cite{Tien:1971,Lacey:1990,Ladouceur:1994,Vlasov:2004,Poulton:2006,Melati:2014}, a recent conclusion being that while propagation losses can certainly arise from material absorption and radiation associated with tight bends, the most significant source of loss in modern integrated waveguides is often scattering due to sidewall roughness inherent in fabrication
processes~\cite{Melati:2014}.

Modern quantum-theoretical treatments of spontaneous photon generation typically proceed along one of two directions. In one school, the focus is on operator expectation values, and differential equations for these operators are developed in analogy with classical coupled mode equations (see e.g.~\cite{Voss:2006,Lin:2007,Silva:2013}). In the other, more of a focus is placed on quantum states (see e.g.~\cite{Grice:1997,Wong:2006,Lvovsky:2007,Yang:2008,Brainis:2009,Christ:2013}). Although the first approach has initially proven more amenable to extensions to include loss~\cite{Voss:2006,Silva:2013}, there is no reason to believe that such a task is not possible in the second.  Indeed, here we extend a multiple-frequency mode Hamiltonian formalism of spontaneous parametric downconversion (SPDC) in waveguides~\cite{Yang:2008} to include scattering loss in the nonlinear region.  

With an eye toward future calculations, we see four key advantages to employing this ``backward Heisenberg picture'' approach, so-named because it evolves operators backward in time to ensure that their associated Schr\"odinger picture states correctly evolve forward in time.  In particular, it works within a wavevector-time framework, rather than frequency-time or position-time, enabling extensions beyond effectively one-dimensional devices to two-dimensional and three-dimensional structures.  Secondly, it correctly accounts for both material and modal dispersion, not only in calculating the phase matching of the process, but also in the normalization of the modes involved~\cite{Bhat:2006}.  Thirdly, the pump pulse is treated fully quantum mechanically, allowing the description of nonlinear quantum optical processes for arbitrary input states of light.  Finally, it  places classical and quantum wave mixing processes within a consistent theoretical framework, making it easy to draw comparisons and develop new physical insights~\cite{Helt:2012,Liscidini:2013}. 

We limit ourselves here to consideration of photon pair generation via SPDC, as this allows us to consider pump losses separately from generated photon losses; nonetheless, we expect many of the results presented here to carry over to photon pair generation via spontaneous four-wave mixing, a topic we intend to explore in detail in future work. While SPDC pair generation in waveguides has been studied in the past~\cite{Yariv:1975,Banaszek:2001,Tanzilli:2002,Zhang:2007,Eckstein:2008,Horn:2013}, the effects of scattering loss have rarely been included explicitly~\cite{Caves:1987,Antonosyan:2014}, and never within a multiple-frequency mode quantum state picture. Indeed, in the analysis of experimental results such loss is usually lumped in with detector efficiencies and losses associated with coupling on and off the chip~\cite{Tanzilli:2001,Zhang:2007}.  Typical theoretical analyses of the utility of photonic states when losses are involved often model loss with the inclusion of asymmetric beam splitters~\cite{Loudon:2000}, or frequency-dependent beam splitters acting as spectral filters~\cite{Branczyk:2010}, placed after the nonlinear region where the photons are generated~\cite{Huver:2008,Dorner:2009,Branczyk:2010,Zhang:2013}.  It is clear that both of these approaches miss any effects due to the \emph{simultaneous} action of nonlinearity and loss, as opposed to effects arising from their \emph{successive} action.

As we show, our approach correctly captures the full spectral structure of SPDC generated photons, including the effects of loss on photon frequency correlations, and enables a prediction of the quantum performance of nonlinear waveguides in the presence of loss. In particular, we show that the common expression for the biphoton probability density, or joint spectral intensity (JSI), in which it is composed of just a pump pulse spectrum term and a phase matching term, should also contain a loss matching term that can strongly modify its shape. We also show that the standard practice of quantifying device performance only in terms of photon pair exit probability should be complemented by specifying the probability of accidental singles, from photon pairs that have lost one photon, exiting the waveguide. Put together, a picture of the trade-offs between frequency separability and photon pair to single photon exit probability emerges. Additionally, as we begin with a coherent state pump in the Schr\"odinger picture and follow its evolution through the device, our approach remains relatively straightforward and can easily be generalized to more complicated input states, additional nonlinear effects~\cite{Helt:2013,Husko:2013}, and various integrated nonlinear structures beyond channel waveguides~\cite{Liscidini:2012}. For a treatment of the simpler problem of including the effects of scattering loss {\it following} photon generation within our formalism, we refer the reader to Helt~\cite{Helt:2013thesis}.

In Section \ref{sec:formalism} we introduce the general formalism, first reviewing how a calculation proceeds in the absence of loss, and then turning to differences that arise when scattering loss is included. In Section \ref{sec:density}, working in the negligible multi-pair generation regime, we construct the reduced density operator associated with at most a single pair of photons exiting the lossy waveguide without scattering, first in wavevector-space and then switching to a frequency representation as well. This density operator is seen to naturally separate into the sum of a two-photon, a single-photon, and a vacuum density operator. In Section \ref{sec:BWF}, still in frequency space, we demonstrate the utility of our derived expressions. We first use the natural splitting of the total reduced density operator to calculate the probabilities with which two photons, one photon, or zero photons exit the waveguide without scattering. We then compare the form of a general lossy biphoton wave function with one calculated in the absence of loss, and note the appearance of both an exponential decay term and a loss matching term. Finally, we consider biphoton probability densities in the three qualitatively different regimes: one in which either the loss of pump photons or generated photons is much higher than the other, one in which the loss of pump photons and generated photons are appropriately balanced, and one in which the loss of generated photons is frequency dependent.  For concreteness we calculate biphoton probability densities as well as the probabilities with which photon pairs and single photons exit the nonlinear device for a realistic Bragg reflection waveguide~\cite{Zhukovsky:2012}. We conclude in Section \ref{sec:conclusions}.

\section{From input pulses to output photons}
\label{sec:formalism}

\subsection{Summary of our formalism in the absence of scattering loss}

The formalism that we extend here was initially presented in an earlier
work~\cite{Yang:2008}, and so we direct the reader there for additional
details, providing just a summary here. It begins with linear and
nonlinear Hamiltonians, which are built up from correctly normalized
expansions of the full electric displacement and magnetic field operators
in terms of the modes of interest of the linear problem. In particular,
we assume that modes labelled by $m=\text{D}$ for downconverted (fundamental) and
$m=\text{P}$ for pump (second harmonic) have been found, and write
\begin{equation}
H_{\text{L}}=\sum_{m=\text{D},\text{P}}\int\text{d}k\,\hbar\omega_{mk}a_{mk}^{\dagger}a_{mk},\label{eq:HL}
\end{equation}
\begin{equation}
H_{\text{NL}}=-\int\text{d}k_{1}\text{d}k_{2}\text{d}k\, S\left(k_{1},k_{2},k\right)a_{\text{D}k_{1}}^{\dagger}a_{\text{D}k_{2}}^{\dagger}a_{\text{P}k}+\text{H.c.},\label{eq:H_NL}
\end{equation}
where H.c. denotes Hermitian conjugate, and
\begin{align}
\left[a_{mk},a_{m^{\prime}k^{\prime}}^{\dagger}\right]= & \delta_{mm^{\prime}}\delta\left(k-k^{\prime}\right),\label{eq:aCom}
\end{align}
with all other commutators evaluating to zero. For simplicity, we have assumed that all generated photons are labelled by D, i.e. we have assumed type-I SPDC, though we note that generalizations are straightforward~\cite{Helt:2009}. We assume that all pump and generated photons travel in the forward (positive $k$) direction, a valid approximation for typical dispersion relations~\cite{Yang:2008}, and therefore here and throughout all wavevector integrals are taken over the positive real axis.  All of the nonlinear
optics lives in the coupling term~\cite{Yang:2008}
\begin{align}
S(k_{1},k_{2},k)=&\sqrt{\frac{\hbar\omega_{\text{D}k_{1}}\hbar\omega_{\text{D}k_{2}}\hbar\omega_{\text{P}k}}{\left(4\pi\right)^{3}\varepsilon_{0}}}\nonumber \\
&\times\frac{\bar{\chi}_{2}L\sinc\left[\left(k_1+k_2-k\right)L/2\right]}{\bar{n}^{3}\sqrt{\mathcal{A}\left(k_{1},k_{2},k\right)}},\label{eq:S}
\end{align}
where the nonlinearity has been assumed to exist between $z=-L/2$ and $z=L/2$, and $\overline{n}$ and $\overline{\chi}_{2}$ are, respectively,
a typical effective index and second-order optical nonlinearity introduced
solely for convenience. In particular, our final results depend on
neither $\overline{n}$ nor $\overline{\chi}_{2}$, as they cancel with
counterparts in the definition of the effective area~\cite{Yang:2008}
\begin{align}
&\mathcal{A}\left(k_{1},k_{2},k\right)=\left|\int_{-\infty}^\infty\text{d}x\text{d}y\right.\nonumber\\
&\quad\times\left.\frac{\bar{n}^{3}\chi_{2}^{ijk}d_{\text{D}k_{1}}^{i}\left(x,y\right)d_{\text{D}k_{2}}^{j}\left(x,y\right)\left[d_{\text{P}k}^{k}\left(x,y\right)\right]^{*}}{\bar{\chi}_{2}\varepsilon_{0}^{3/2}n^{2}\left(x,y;\omega_{\text{D}k_{1}}\right)n^{2}\left(x,y;\omega_{\text{D}k_{2}}\right)n^{2}\left(x,y;\omega_{\text{P}k}\right)}\right|^{-2},
\end{align}
with $d_{mk}^{i}\left(x,y\right)$ the $i$-th component of the displacement
field at wavenumber $k$, and $n\left(x,y;\omega_{mk}\right)$ the
material refractive index at wavenumber $k$, both at waveguide cross-sectional
position $(x,y)$. We have chosen the field amplitudes such that we
can take the phase associated with the effective area to be zero.

We frame evolution through the nonlinear waveguide in terms of `asymptotic-in'
and `-out' states, borrowing from scattering theory. Their introduction
eliminates trivial linear evolution from our main calculation, as
the asymptotic-in state is defined as the state evolved from $t=t_{0}$,
with energy localized at the beginning of the waveguide, to $t=0$,
at its centre, according to only $H_{\text{L}}$. Similarly, the asymptotic-out
state is defined as the state at $t=0$ that would evolve to $t=t_{1}$,
with energy localized at the end of the waveguide, if the evolution
occurred according to the same linear Hamiltonian. The duration of
the interaction is on the order of the length of the assumed nonlinear portion of waveguide, $L$,
divided by the group velocity of the pump field, $v_{\text{P}}$,
i.e. $t_{1}-t_{0}\approx L/v_{\text{P}}$. However, as in scattering theory,
it is common to take $t_{0}\rightarrow-\infty$, $t_{1}\rightarrow\infty$.
The calculation seeks the state of generated photons for an asymptotic-in
coherent state
\begin{equation}
\left\vert \psi\right\rangle _{\text{in}}=\exp\left(z\int\text{d}k\,\phi_{\text{P}}\left(k\right)a_{\text{P}k}^{\dagger}-\text{H.c.}\right)\left\vert \text{vac}\right\rangle ,\label{eq:in}
\end{equation}
with $\left\vert z\right\vert ^{2}$ the average number of photons
per pulse for a normalized pump pulse waveform $\phi_{\text{P}}\left(k\right)$
and $\left\vert \text{vac}\right\rangle =\left\vert \text{vac}\right\rangle _{\text{D}}\otimes\left\vert \text{vac}\right\rangle _{\text{P}}$.
In particular, we proceed by solving for the associated asymptotic-out
state
\begin{equation}
\left\vert \psi\right\rangle _{\text{out}}=e^{iH_{\text{L}}t_{1}/\hbar}e^{-i\left(H_{\text{L}}+H_{\text{NL}}\right)\left(t_{1}-t_{0}\right)/\hbar}e^{-iH_{\text{L}}t_{0}/\hbar}\left\vert \psi\right\rangle _{\text{in}},\label{eq:LinEv}
\end{equation}
in the backward Heisenberg picture~\cite{Yang:2008}, and seek an
output state of the form 
\begin{equation}
\left\vert \psi\right\rangle _{\text{out}}=\exp\left(z\int\text{d}k\,\phi_{\text{P}}\left(k\right)\bar{a}_{\text{P}k}^{\dagger}\left(t_{0}\right)-\text{H.c.}\right)\left\vert \text{vac}\right\rangle,
\end{equation}
where $\bar{a}_{\text{P}k}^\dagger\left(t\right)$ contains a contribution from the D  mode operators. This form follows from our intuition that, for an undepleted
pump, the P mode will remain in a coherent state. Furthermore, any
difference between $\bar{a}_{\text{P}k}^{\dagger}\left(t_{0}\right)$
and $a_{\text{P}k}^{\dagger}$ can be identified with photon creation
and, to first order, is naturally represented as a squeezing operation~\cite{Yang:2008}.
For a general barred operator $\overline{O}\left(t\right)$, which
is seen to evolve backward in time as $\left\vert \psi\right\rangle _{\text{in}}$
evolves forward in time, one can show 
\begin{equation}
i\hbar\frac{\text{d}\overline{O}\left(t\right)}{\text{d}t}=\left[\overline{O}\left(t\right),V\left(t\right)\right],\label{eq:Heisenberg}
\end{equation}
subject to the ``final'' condition 
\begin{equation}
\overline{O}\left(t_{1}\right)=O,\label{eq:final_condition}
\end{equation}
with 
\begin{align}
V\left(t\right)=&-\int\text{d}k_{1}\text{d}k_{2}\text{d}k\, S\left(k_{1},k_{2},k;t\right)\overline{a}_{\text{D}k_{1}}^{\dagger}\left(t\right)\overline{a}_{\text{D}k_{2}}^{\dagger}\left(t\right)\overline{a}_{\text{P}k}\left(t\right)\nonumber \\
&+\text{H.c.},
\end{align}
and $S\left(k_{1},k_{2},k;t\right)=S\left(k_{1},k_{2},k\right)e^{i\left(\omega_{\text{D}k_{1}}+\omega_{\text{D}k_{2}}-\omega_{\text{P}k}\right)t}$.
In short, the problem of calculating the evolution of a state through
the waveguide reduces to the integration of \eqref{eq:Heisenberg}
from $t=t_{1}$ back to $t=t_{0}$ subject to \eqref{eq:final_condition}
or solving differential equations for barred (backward Heisenberg)
operators. In the undepleted pump approximation, the approximate solution is~\cite{Yang:2008}
\begin{equation}
\left\vert \psi\right\rangle _{\text{out}}=\exp\left(\zeta C_\text{II}^{\dagger}\left(t_{0}\right)-\text{H.c.}\right)\left\vert \text{vac}\right\rangle _{\text{D}}\otimes\left\vert \psi\right\rangle _{\text{P}},\label{eq:out}
\end{equation}
where $\left\vert \psi\right\rangle _{\text{P}}=\exp\left(z\int\text{d}k\,\phi_{\text{P}}\left(k\right)a_{\text{P}k}^{\dagger}-\text{H.c.}\right)\left\vert \text{vac}\right\rangle _{\text{P}}$ is the input coherent state, and
\begin{equation}
C_\text{II}^{\dagger}\left(t_{0}\right)=\frac{1}{\sqrt{2}}\int\text{d}k_{1}\text{d}k_{2}\,\phi\left(k_{1},k_{2}\right)a_{\text{D}k_{1}}^{\dagger}a_{\text{D}k_{2}}^{\dagger},\label{eq:CiiNoLoss}
\end{equation}
is a two-photon creation operator characterized by the biphoton wave
function
\begin{equation}
\phi\left(k_{1},k_{2}\right)=\frac{\sqrt{2}z}{\zeta}\frac{i}{\hbar}\int\text{d}k\,\phi_{\text{P}}\left(k\right)\int_{t_{0}}^{t_{1}}\text{d}\tau\, S\left(k_{1},k_{2},k;\tau\right).\label{eq:PhiNoLoss}
\end{equation}
Note that it is symmetric 
\begin{equation}
\phi\left(k_{1},k_{2}\right)=\phi\left(k_{2},k_{1}\right).
\end{equation}
Although \eqref{eq:out} is already normalized, we are also free to choose $C_\text{II}^{\dagger}\left(t_{0}\right)\left\vert \text{vac}\right\rangle _{\text{D}}$
to be normalized, which requires that we set
\begin{equation}
\int\text{d}k_{1}\text{d}k_{2}\left|\phi\left(k_{1},k_{2}\right)\right|^{2}=1.
\end{equation}
Recalling \eqref{eq:out}, we see that this choice implies that
in the limit $\left\vert \zeta\right\vert \ll1$, $\left\vert \zeta\right\vert ^{2}$
can be thought of as the average number of generated photon pairs
per pump pulse.

\subsection{Photon generation in the presence of scattering loss}

To include the effects of scattering loss in nonlinear waveguides
within this formalism, we introduce a reservoir of radiation
modes with the free Hamiltonian
\begin{equation}
H_{\text{R}}=\sum_{m=\text{D},\text{P}}\int\text{d}k\text{d}\mu\,\hbar\Omega_{m\mu k}b_{m\mu k}^{\dagger}b_{m\mu k},\label{eq:HR}
\end{equation}
containing operators that satisfy 
\begin{equation}
\left[b_{m\mu k},b_{m\mu^{\prime}k^{\prime}}^{\dagger}\right]=\delta_{mm^{\prime}}\delta\left(\mu-\mu^{\prime}\right)\delta\left(k-k^{\prime}\right),\label{eq:commutators}
\end{equation}
and which are coupled to pump and downconverted modes via the Hamiltonian
\begin{equation}
H_{\text{C}}=\sum_{m=\text{D},\text{P}}\int\text{d}k\text{d}\mu\,\hbar\left(c_{m\mu k}a_{mk}^{\dagger}b_{m\mu k}+c_{m\mu k}^{\ast}b_{m\mu k}^{\dagger}a_{mk}\right).\label{eq:HC}
\end{equation}
Here $\mu$ is a shorthand for all quantities necessary to specify
reservoir modes in addition to $m$ and $k$, and $c_{m\mu k}$ are
waveguide-reservoir coupling terms. While it is possible in principle
to solve for the radiation modes with which the $b_{m\mu k}$ are associated, and perform field overlap integrals to determine the $c_{m\mu k}$,
we do not attempt to do so here. We simply view the reservoir and
coupling Hamiltonians as phenomenological entities, and later on in
our calculation connect the waveguide-reservoir coupling terms to
the usual experimentally determined loss coefficients. Although
$b_{\text{D}\mu k}$ and $b_{\text{P}\mu k}$ come from the same
field expansion, the division of the ``total'' reservoir into two
parts through the sums in \eqref{eq:HR} and \eqref{eq:HC} is nevertheless
justified. Following the rotating wave approximation used to simplify
the coupling Hamiltonian, $H_{\text{C}}$, the reservoir modes that
couple strongly to the D waveguide modes, $a_{\text{D}k}$, are well-separated
in frequency from the reservoir modes that couple strongly to the
P waveguide modes, $a_{\text{P}k}$. While this phenomenological model
allows for light to couple both into and out of the guided waveguide modes via the reservoir, we eventually take the temperature of the reservoir to
be zero so that no light can ever scatter back into the waveguide
modes. This assumption, which is quite reasonable here
as room temperature blackbody radiation at the frequencies of interest
is negligible, is common in the study of open quantum optical systems and simplifies
our calculations.

At this point, there are many ways that one could proceed. One could
derive an expression for the evolution of the reduced density operator
describing the generated photons. Written as a differential equation,
this is known as the Master Equation, and can be put into Lindblad
form \cite{Lindblad:1976}. Alternatively, one could work with quasiprobability
distribution function representations for the same density operator~\cite{Carmichael:1999},
and arrive at a Fokker-Planck type equation \cite{Fokker:1914,Haken:1981}.
However, we find that the approach that most clearly captures the
physics and lends itself to integration with a Schr\"odinger state picture
approach is a quantum Langevin formalism in which equations of motion
are derived for waveguide operators in terms of ``fluctuating force''
reservoir operators, which are later traced out of the appropriate
density operator \cite{Haken:1981}, and it is this procedure that
we follow here.

We imagine the same generalized asymptotic-in coherent state as above
\eqref{eq:in} incident on a lossy nonlinear waveguide, and also work
in the backward Heisenberg picture as above, the only difference
being that we now include the reservoir and coupling Hamiltonians
in our calculation. With $H_{\text{L}}\rightarrow H_{\text{L}}+H_{\text{R}}$
in \eqref{eq:LinEv}, we find that the asymptotic-out state can be written
\begin{equation}
\left\vert \psi\right\rangle _{\text{out}}=\exp\left(z\int\text{d}k\,\phi_{\text{P}}\left(k\right)\overline{a}_{\text{P}k}^{\dagger}\left(t_{0}\right)-\text{H.c.}\right)\left\vert \text{vac}\right\rangle ,\label{eq:out_new}
\end{equation}
where barred operators $\overline{O}\left(t\right)$ now evolve according
to the backward Heisenberg equation 
\begin{equation}
i\hbar\frac{\text{d}\overline{O}\left(t\right)}{\text{d}t}=\left[\overline{O}\left(t\right),W\left(t\right)\right],\label{eq:BackwardHeisenberg}
\end{equation}
with 
\begin{align}
W&\left(t\right)\nonumber \\
= & \sum_{m=\text{D},\text{P}}\int\text{d}k\text{d}\mu\,\hbar\left(c_{m\mu k}\overline{a}_{mk}^{\dagger}\left(t\right)\overline{b}_{m\mu k}\left(t\right)e^{i\left(\omega_{mk}-\Omega_{m\mu k}\right)t}+\text{H.c.}\right)\nonumber \\
 & -\int\text{d}k_{1}\text{d}k_{2}\text{d}k\, S\left(k_{1},k_{2},k;t\right)\overline{a}_{\text{D}k_{1}}^{\dagger}\left(t\right)\overline{a}_{\text{D}k_{2}}^{\dagger}\left(t\right)\overline{a}_{\text{P}k}\left(t\right)+\text{H.c.,}
\end{align}
subject to the ``final'' condition 
\begin{equation}
\overline{O}\left(t_{1}\right)=O.
\end{equation}
The vacuum ket now encompasses the Hilbert space of the reservoir in addition to the D and P waveguide operator spaces, i.e. $\left\vert \text{vac}\right\rangle =\left\vert \text{vac}\right\rangle _{\text{D}}\otimes\left\vert \text{vac}\right\rangle _{\text{P}}\otimes\left\vert \text{vac}\right\rangle _{\text{R}}$. Writing the state of the reservoir in this way is justified in the zero-temperature limit, where the reservoir density operator 
\begin{equation}
\rho_\text{R}\left(T\right) = \frac{e^{-H_\text{R}/\left(k_\text{B} T\right)}}{\Tr_\text{R}\left[e^{-H_\text{R}/\left(k_\text{B} T\right)}\right]},
\end{equation}
with $T$ the temperature and $k_\text{B}$ the Boltzmann constant, becomes the pure state 
\begin{equation}
\rho_\text{R}\left(0\right) = \left\vert\text{vac}\right\rangle_\text{R}\left\langle\text{vac}\right\vert_\text{R}\equiv\rho_\text{R}.
\label{eq:rhoR}
\end{equation}

The method presented here has eliminated trivial linear evolution according to both
$H_{\text{L}}$ as well as $H_{\text{R}}$ and, in addition to nonlinear
effects from $H_{\text{NL}}$, our backward Heisenberg equation
now includes the effects of coupling to the reservoir. Explicitly,
the differential equations for the barred reservoir operators, which
follow from \eqref{eq:BackwardHeisenberg}, are 
\begin{equation}
i\hbar\frac{\text{d}\overline{b}_{m\mu k}^{\dagger}\left(t\right)}{\text{d}t}=-\hbar c_{m\mu k}\overline{a}_{mk}^{\dagger}\left(t\right)e^{i\left(\omega_{mk}-\Omega_{m\mu k}t\right)}.
\end{equation}
They are formally solved as 
\begin{equation}
\overline{b}_{m\mu k}^{\dagger}\left(t\right)=b_{m\mu k}^{\dagger}-i\int_{t}^{t_{1}}c_{m\mu k}\overline{a}_{mk}^{\dagger}\left(\tau\right)e^{i\left(\omega_{mk}-\Omega_{m\mu k}t\right)}\text{d}\tau,\label{eq:bSol}
\end{equation}
which can then be substituted in the differential equations
for the barred waveguide operators $\overline{a}_{mk}\left(t\right)$.
Treating the nonlinear term containing $S\left(k_{1},k_{2},k;t\right)$
as a perturbation, we are interested in the first-order solution for
the waveguide operator in our output state $\overline{a}_{\text{P}k}\left(t\right)$.
This operator has the first-order equation 
\begin{align}
\frac{\text{d}}{\text{d}t} & \left(\overline{a}_{\text{P}k}^{\dagger}\left(t\right)\right)^{1}\nonumber \\
= & \int\text{d}\mu\left(\vphantom{\left(\overline{a}_{\text{P}k}^{\dagger}\left(t\right)\right)^{1}}ib_{\text{P}\mu k}^{\dagger}c_{\text{P}\mu k}^{\ast}e^{-i\left(\omega_{\text{P}k}-\Omega_{\text{P}\mu k}\right)t}\right.\nonumber \\
&+\left.\int_t^{t_1}\text{d}\tau\left(\overline{a}_{\text{P}k}^{\dagger}\left(\tau\right)\right)^{1}\left\vert c_{\text{P}\mu k}\right\vert ^{2}e^{i\left(\omega_{\text{P}k}-\Omega_{\text{P}\mu k}\right)\left(\tau-t\right)}\right)\nonumber \\
 & +\frac{1}{i\hbar}\int\text{d}k_{1}\text{d}k_{2}S\left(k_{1},k_{2},k;t\right)\left(\overline{a}_{\text{D}k_{1}}^{\dagger}\left(t\right)\right)^{0}\left(\overline{a}_{\text{D}k_{2}}^{\dagger}\left(t\right)\right)^{0},\label{eq:first}
\end{align}
with zeroth-order equations for both D and P barred waveguide operators
\begin{align}
\frac{\text{d}}{\text{d}t} & \left(\overline{a}_{mk}^{\dagger}\left(t\right)\right)^{0}\nonumber \\
= & \int\text{d}\mu\left(\vphantom{\left(\overline{a}_{mk}^{\dagger}\left(t\right)\right)^{0}}ib_{m\mu k}^{\dagger}c_{m\mu k}^{\ast}e^{-i\left(\omega_{mk}-\Omega_{m\mu k}\right)t}\right.\nonumber \\
 & +\left.\int_{t}^{t_{1}}\text{d}\tau\left(\overline{a}_{mk}^{\dagger}\left(\tau\right)\right)^{0}\left\vert c_{m\mu k}\right\vert ^{2}e^{i\left(\omega_{mk}-\Omega_{m\mu k}\right)\left(\tau-t\right)}\right),\label{eq:zeroth}
\end{align}
where we have used \eqref{eq:bSol}. We remind readers that $c_{m\mu k}$
are coupling coefficients, not operators, and that $\hbar\omega_{mk}$
and $\hbar\Omega_{m\mu k}$ are, respectively, energies associated
with waveguide mode and radiation mode (reservoir) operators.

To evaluate these integrals we note that, as a first approximation,
the guided modes labelled by a particular $m$ and $k$ couple equally
to all reservoir modes of the same $m$ and $k$ regardless of $\mu$.
That is, for a fixed $m$ and $k$, there are so many degrees of freedom
represented by $\mu$ that scattering into each is equally likely.
Physically, this is because we assume that waveguide roughness at different positions is
approximately uncorrelated and the scattering spectrum is flat over
the $\mu$ of interest, an approximation certainly accurate down to
a few nanometres \cite{Melati:2014}. With this in mind, we approximate
$\left\vert c_{m\mu k}\right\vert ^{2}\approx \mathcal{C}_{mk}$ as independent
of $\mu$ and write $\text{d}\mu=\left(\text{d}\mu/\text{d}\Omega_{m\mu k}\right)\text{d}\Omega_{m\mu k}$,
also approximating the density of states $\text{d}\mu/\text{d}\Omega_{m\mu k}\approx\mathcal{D}_{mk}$
as independent of $\mu$ so that \eqref{eq:zeroth} can be cast into
the form of a quantum mechanical Langevin equation 
\begin{equation}
\frac{\text{d}}{\text{d}t}\left(\overline{a}_{mk}^{\dagger}\left(t\right)\right)^{0}=F_{mk}^{\dagger}\left(t\right)+\beta_{mk}\left(\overline{a}_{mk}^{\dagger}\left(t\right)\right)^{0},
\label{eq:QLange}
\end{equation}
where the loss rate $\beta_{mk}=\mathcal{C}_{mk}\pi\mathcal{D}_{mk}>0$ and
fluctuating force operator
\begin{equation}
F_{mk}^{\dagger}\left(t\right)=i\int\text{d}\mu\, b_{m\mu k}^{\dagger}c_{m\mu k}^{\ast}e^{-i\left(\omega_{mk}-\Omega_{m\mu k}\right)t}.
\end{equation}
The zeroth-order solution for $\overline{a}_{mk}^{\dagger}\left(t\right)$
is then 
\begin{equation}
\left(\overline{a}_{mk}^{\dagger}\left(t\right)\right)^{0}=a_{mk}^{\dagger}e^{-\beta_{mk}\left(t_{1}-t\right)}-\int_{t}^{t_{1}}\text{d}\tau\, F_{mk}^{\dagger}\left(\tau\right)e^{-\beta_{mk}\left(\tau-t\right)}.
\end{equation}
It is easy to verify that the fluctuating force operators satisfy (recall
\eqref{eq:commutators}) 
\begin{equation}
\left[F_{mk}\left(t\right),F_{m^{\prime}k^{\prime}}^{\dagger}\left(t^{\prime}\right)\right]=2\beta_{mk}\delta_{mm^{\prime}}\delta\left(t-t^{\prime}\right)\delta\left(k-k^{\prime}\right),\label{eq:FComm}
\end{equation}
and thus that the equal time commutation relation for $\left(\overline{a}_{mk}^{\dagger}\left(t\right)\right)^{0}$
does not decay in time, as quantum mechanics requires. In fact,
the commutation relation is the same as that for $a_{mk}^{\dagger}$
\eqref{eq:aCom}. We remark that these zeroth-order equations are just
multimode generalizations of the well-known single-mode quantum Langevin
equation \cite{Haken:1981}, $\text{d}a^{\dagger}/\text{d}t=-\beta a^{\dagger}+F^{\dagger}\left(t\right)$, with $\left[F\left(t\right),F^{\dagger}\left(t^{\prime}\right)\right]=2\beta\delta\left(t-t^{\prime}\right)$. The difference in sign of the loss rate $\beta$ between the two differential equations (recall \eqref{eq:QLange}) arises because in the usual Heisenberg picture system operators evolve forward in time, whereas in the backward Heisenberg picture here the system operators evolve backward in time such that their associated Schr\"odinger picture state correctly evolves forward in time.
The first-order solution for $\overline{a}_{\text{P}k}^{\dagger}\left(t\right)$
follows immediately from \eqref{eq:first}, and we see that to this
order the asymptotic-out state \eqref{eq:out_new} can be written
\begin{equation}
\left\vert \psi\right\rangle _{\text{out}}=\exp\left(\zeta C_\text{II}^{\dagger}\left(t_{0}\right)-\text{H.c.}\right)\left\vert \text{vac}\right\rangle _{\text{D}}\otimes\left\vert \text{vac}\right\rangle _{\text{R}}\otimes\left\vert \psi\right\rangle _{\text{P}}\label{eq:out2}
\end{equation}
where
\begin{align}
&\left\vert \psi\right\rangle _{\text{P}}=\exp\left[z\int\text{d}k\,\phi_{\text{P}}\left(k\right)\right.\nonumber\\
&\times\left.\left(a_{\text{P}k}^{\dagger}e^{-\beta_{\text{P}k}\left(t_{1}-t_{0}\right)}-\int_{t_0}^{t_1}\text{d}\tau F_{\text{P}k}^\dagger\left(\tau\right)e^{-\beta_{\text{P}k}\left(\tau-t_0\right)}\right)-\text{H.c.}\right]\left\vert \text{vac}\right\rangle _{\text{P}},
\end{align}
is, after tracing over the fluctuating force operator, the initial coherent state \eqref{eq:in}, having decayed exponentially
in time from $t_{0}$ to $t_{1}$ and 
\begin{align}
C_\text{II}^{\dagger}\left(t_0\right)= & \frac{1}{\sqrt{2}}\int_{t_0}^{t_{1}}\text{d}\tau\int\text{d}k_{1}\text{d}k_{2}\,\phi\left(k_{1},k_{2};\tau\right)\nonumber \\
 & \times\left(a_{\text{D}k_{1}}^{\dagger}e^{-\beta_{\text{D}k_{1}}\left(t_{1}-\tau\right)}-\int_{\tau}^{t_{1}}\text{d}\tau^{\prime}F_{\text{D}k_{1}}^{\dagger}\left(\tau^{\prime}\right)e^{-\beta_{\text{D}k_{1}}\left(\tau^{\prime}-\tau\right)}\right)\nonumber \\
 & \times\left(a_{\text{D}k_{2}}^{\dagger}e^{-\beta_{\text{D}k_{2}}\left(t_{1}-\tau\right)}-\int_{\tau}^{t_{1}}\text{d}\tau^{\prime}F_{\text{D}k_{2}}^{\dagger}\left(\tau^{\prime}\right)e^{-\beta_{\text{D}k_{2}}\left(\tau^{\prime}-\tau\right)}\right),\label{eq:CII}
\end{align}
is a two-photon creation operator characterized by the
total biphoton wave function 
\begin{equation}
\phi\left(k_{1},k_{2};\tau\right)=\frac{\sqrt{2}z}{\zeta}\frac{i}{\hbar}\int\text{d}k\,\phi_{\text{P}}\left(k\right)S\left(k_{1},k_{2},k;\tau\right)e^{-\beta_{\text{P}k}\left(\tau-t_0\right)}.\label{eq:firstphi}
\end{equation}
The two-photon creation operator and biphoton wave function calculated in the presence
of loss are key results. They represent generalizations of \eqref{eq:CiiNoLoss} and \eqref{eq:PhiNoLoss}
to include the effects of scattering
loss for photons generated in nonlinear waveguides. However, much
more can be learned in the frequency representation, and although
we have progressed from the $c_{m\mu k}$ to $\beta_{mk}$, the $\beta_{mk}$
are still not in the form of the standard attenuation coefficients $\alpha_{mk}$, usually expressed as inverse lengths.

Note that, just as seen for its non-lossy counterpart above \eqref{eq:PhiNoLoss},
the biphoton wave function here \eqref{eq:firstphi} is symmetric
\begin{equation}
\phi\left(k_{1},k_{2};\tau\right)=\phi\left(k_{2},k_{1};\tau\right).\label{eq:sym}
\end{equation}
Also as above, although \eqref{eq:out2} is normalized, we are free to choose $C_\text{II}^{\dagger}\left(t_{0}\right)\left\vert \text{vac}\right\rangle _{\text{D}}$
to be normalized, implying that in the limit $\left\vert \zeta\right\vert \ll1$, $\left\vert \zeta\right\vert ^{2}$
can still be thought of as the average number of photon pairs generated by each pump pulse, although now some of the photons could be lost before exiting the device.  We will consider this condition explicitly in the next Section.  For now we  note that if there were no loss, i.e. no coupling to the reservoirs $\left(c_{m\mu k}=0\right)$, then $\beta_{mk}=0$, $F_{mk}\left(t\right)=0$, and \eqref{eq:CII} would be the same as \eqref{eq:CiiNoLoss}.

\section{The density operator representation of generated photon pairs}

\label{sec:density}

We now turn to the statistical properties of the generated photon states by moving to a density operator picture.  We imagine that we are in a pulsed pump regime where the
probability of generating a pair of photons is low enough, $\left\vert \zeta\right\vert \ll1$,
that we may approximate 
\begin{equation}
\left\vert \psi\right\rangle _{\text{out}}=\left(1+\zeta C_\text{II}^{\dagger}\left(t_{0}\right)\right)\left\vert \text{vac}\right\rangle _{\text{D}}\otimes\left\vert \text{vac}\right\rangle _{\text{R}}\otimes\left\vert \psi\right\rangle _{\text{P}}.
\end{equation}
Anticipating the eventual trace over reservoir operators, we can write this in density
operator form as
\begin{equation}
\rho_{\text{out}}=\rho_{\text{D}}\otimes\rho_{\text{P}},
\end{equation}
where (recall \eqref{eq:rhoR})
\begin{align}
\rho_{\text{D}}=&\left(\left\vert \text{vac}\right\rangle _{\text{D}}+\zeta C_\text{II}^{\dagger}\left(t_{0}\right)\left\vert \text{vac}\right\rangle _{\text{D}}\right)\rho_{\text{R}}\nonumber \\
&\times\left(\left\langle \text{vac}\right\vert _{\text{D}}+\left\langle \text{vac}\right\vert _{\text{D}}C_\text{II}\left(t_{0}\right)\zeta^{\ast}\vphantom{C_\text{II}^{\dagger}}\right),
\label{eq:rhoF}
\end{align}
and 
\begin{equation}
\rho_{\text{P}}=\left\vert \psi\right\rangle _{\text{P}}\left\langle \psi\right\vert _{\text{P}}.
\end{equation}
While vacuum-pair correlations, or ``cross terms'', in \eqref{eq:rhoF} are present, we note that they would be present even if one constructed a density operator in the low probability of pair production regime in a lossless calculation.  There (recall \eqref{eq:out}) one also has $\left(1+\zeta C_\text{II}^{\dagger}\left(t_{0}\right)\right)\left\vert \text{vac}\right\rangle _{\text{D}}$, and often focuses on the normalized two-photon state $C_\text{II}^{\dagger}\left(t_{0}\right)\left\vert \text{vac}\right\rangle _{\text{D}}$, produced with probability $\left\vert\zeta\right\vert^2$. As our interest in this work is in photon pairs, we focus on the corresponding piece of our density operator
\begin{equation}
\rho_\text{II} \equiv C_\text{II}^{\dagger}\left(t_{0}\right)\left\vert \text{vac}\right\rangle _{\text{D}}\rho_{\text{R}}\left\langle \text{vac}\right\vert _{\text{D}}C_\text{II}\left(t_{0}\right),
\end{equation}
produced with probability $\left\vert\zeta\right\vert^2$, in all that follows.

\subsection{$k$-space expressions}

 We construct the reduced density operator describing
the state of generated photons by tracing over the reservoir operators
in $\rho_\text{II}$ at zero temperature (see Appendix \ref{sec:comms})
\begin{widetext}
\begin{align}
\rho_{\text{gen}}= & \Tr_{\text{R}}^{T=0K}\left[C_\text{II}^{\dagger}\left(t_{0}\right)\left\vert \text{vac}\right\rangle _{\text{D}}\rho_{\text{R}}\left\langle \text{vac}\right\vert _{\text{D}}C_\text{II}\left(t_{0}\right)\right]\nonumber \\
= & \frac{1}{2}\int_{t_{0}}^{t_{1}}\text{d}\tau\text{d}\tau^{\prime}\int\text{d}k_{1}\text{d}k_{2}\text{d}k_{1}^{\prime}\text{d}k_{2}^{\prime}\,\phi\left(k_{1},k_{2};\tau\right)\phi^{\ast}\left(k_{1}^{\prime},k_{2}^{\prime};\tau^{\prime}\right)\nonumber \\
 & \quad\times a_{\text{D}k_{1}}^{\dagger}e^{-\beta_{\text{D}k_{1}}\left(t_{1}-\tau\right)}a_{\text{D}k_{2}}^{\dagger}e^{-\beta_{\text{D}k_{2}}\left(t_{1}-\tau\right)}\left\vert \text{vac}\right\rangle _{\text{D}}\left\langle \text{vac}\right\vert _{\text{D}}a_{\text{D}k_{1}^{\prime}}e^{-\beta_{\text{D}k_{1}^{\prime}}\left(t_{1}-\tau^{\prime}\right)}a_{\text{D}k_{2}^{\prime}}e^{-\beta_{\text{D}k_{2}^{\prime}}\left(t_{1}-\tau^{\prime}\right)}\nonumber \\
 & +2\int_{t_{0}}^{t_{1}}\text{d}\tau\text{d}\tau^{\prime}\int\text{d}k_{1}\text{d}k_{2}\text{d}k^{\prime}\,\phi\left(k_{1},k^{\prime};\tau\right)\phi^{\ast}\left(k_{2},k^{\prime};\tau^{\prime}\right)\left(e^{-\beta_{\text{D}k^{\prime}}\left\vert\tau-\tau^{\prime}\right\vert}-e^{-\beta_{\text{D}k^{\prime}}\left(2t_{1}-\tau-\tau^{\prime}\right)}\right)\nonumber \\
 & \quad\times a_{\text{D}k_{1}}^{\dagger}e^{-\beta_{\text{D}k_{1}}\left(t_{1}-\tau\right)}\left\vert \text{vac}\right\rangle _{\text{D}}\left\langle \text{vac}\right\vert _{\text{D}}a_{\text{D}k_{2}}e^{-\beta_{\text{D}k_{2}}\left(t_{1}-\tau^{\prime}\right)}\nonumber \\
 & +\int_{t_{0}}^{t_{1}}\text{d}\tau\text{d}\tau^{\prime}\int\text{d}k_{1}\text{d}k_{2}\,\phi\left(k_{1},k_{2};\tau\right)\phi^{\ast}\left(k_{1},k_{2};\tau^{\prime}\right)\nonumber \\
 & \quad\times\left(e^{-\beta_{\text{D}k_1}\left\vert\tau-\tau^{\prime}\right\vert}-e^{-\beta_{\text{D}k_{1}}\left(2t_{1}-\tau-\tau^{\prime}\right)}\right)\left(e^{-\beta_{\text{D}k_2}\left\vert\tau-\tau^{\prime}\right\vert}-e^{-\beta_{\text{D}k_{2}}\left(2t_{1}-\tau-\tau^{\prime}\right)}\right)\left\vert \text{vac}\right\rangle _{\text{D}}\left\langle \text{vac}\right\vert _{\text{D}},
\label{eq:trace}
\end{align}
\end{widetext}
where we have used \eqref{eq:sym}. Thus our state of generated photons naturally separates into
\begin{widetext}
\begin{align}
\rho_{2}= & \frac{1}{2}\int_{t_{0}}^{t_{1}}\text{d}\tau\int\text{d}k_{1}\text{d}k_{2}\,\phi\left(k_{1},k_{2};\tau\right)a_{\text{D}k_{1}}^{\dagger}e^{-\beta_{\text{D}k_{1}}\left(t_{1}-\tau\right)}a_{\text{D}k_{2}}^{\dagger}e^{-\beta_{\text{D}k_{2}}\left(t_{1}-\tau\right)}\left\vert \text{vac}\right\rangle \nonumber \\
 & \times\int_{t_{0}}^{t_{1}}\text{d}\tau^{\prime}\int\text{d}k_{1}^{\prime}\text{d}k_{2}^{\prime}\,\phi^{\ast}\left(k_{1}^{\prime},k_{2}^{\prime};\tau^{\prime}\right)\left\langle \text{vac}\right\vert a_{\text{D}k_{1}^{\prime}}e^{-\beta_{\text{D}k_{1}^{\prime}}\left(t_{1}-\tau^{\prime}\right)}a_{\text{D}k_{2}^{\prime}}e^{-\beta_{\text{D}k_{2}^{\prime}}\left(t_{1}-\tau^{\prime}\right)},\label{eq:pk2}\\
\rho_{1}= & 2\int_{t_{0}}^{t_{1}}\text{d}\tau\text{d}\tau^{\prime}\int\text{d}k_{1}\text{d}k_{2}\text{d}k^{\prime}\,\phi\left(k_{1},k^{\prime};\tau\right)\phi^{\ast}\left(k_{2},k^{\prime};\tau^{\prime}\right)\left(e^{-\beta_{\text{D}k^{\prime}}\left\vert\tau-\tau^{\prime}\right\vert}-e^{-\beta_{\text{D}k^{\prime}}\left(2t_{1}-\tau-\tau^{\prime}\right)}\right)\nonumber \\
 & \times a_{\text{D}k_{1}}^{\dagger}e^{-\beta_{\text{D}k_{1}}\left(t_{1}-\tau\right)}\left\vert \text{vac}\right\rangle \left\langle \text{vac}\right\vert a_{\text{D}k_{2}}e^{-\beta_{\text{D}k_{2}}\left(t_{1}-\tau^{\prime}\right)},\label{eq:pk1}\\
\rho_{0}= & \int_{t_{0}}^{t_{1}}\text{d}\tau\text{d}\tau^{\prime}\int\text{d}k_{1}\text{d}k_{2}\,\phi\left(k_{1},k_{2};\tau\right)\phi^{\ast}\left(k_{1},k_{2};\tau^{\prime}\right)\nonumber \\
 & \times\left(e^{-\beta_{\text{D}k_1}\left\vert\tau-\tau^{\prime}\right\vert}-e^{-\beta_{\text{D}k_{1}}\left(2t_{1}-\tau-\tau^{\prime}\right)}\right)\left(e^{-\beta_{\text{D}k_2}\left\vert\tau-\tau^{\prime}\right\vert}-e^{-\beta_{\text{D}k_{2}}\left(2t_{1}-\tau-\tau^{\prime}\right)}\right)\left\vert \text{vac}\right\rangle \left\langle \text{vac}\right\vert ,\label{eq:pk0}
\end{align}
\end{widetext}
where we have dropped the mode label D on the vacuum state for notational
convenience, leaving
\begin{equation}
\rho_{\text{gen}}=\rho_{2}+\rho_{1}+\rho_{0}.\label{eq:rhogenk}
\end{equation}
As $\rho_{2}$ factors into a ket and a bra, we recognize it as representing a pure state of two photons
with photon creation operators multiplied by decaying exponentials
and correlations determined by $\phi\left(k_{1},k_{2};\tau\right)$.
However the middle term, $\rho_{1}$, can not be factored, and represents a mixed state
of single photons. The final term, $\rho_{0}$, is simply a number
that sits in front of the vacuum density operator $\left\vert \text{vac}\right\rangle \left\langle \text{vac}\right\vert $.
Written in terms of these three pieces, our choice that $C_\text{II}^\dagger\left(t_0\right)\left\vert\text{vac}\right\rangle_\text{D}$ be normalized becomes
\begin{align}
\Tr_\text{D}\left[\rho_{\text{gen}}\right]= & \Tr_\text{D}\left[\rho_{2}\right]+\Tr_\text{D}\left[\rho_{1}\right]+\Tr_\text{D}\left[\rho_{0}\right]\nonumber \\
= & \mathcal{P}_{2}+\mathcal{P}_{1}+\mathcal{P}_{0}\nonumber \\
= & 1,
\end{align}
where
\begin{align}
\mathcal{P}_{\text{II}}=&\int\text{d}k_{1}\text{d}k_{2}\int_{t_{0}}^{t_{1}}\text{d}\tau\text{d}\tau^{\prime}\phi\left(k_{1},k_{2};\tau\right)\nonumber\\
&\times\phi^{\ast}\left(k_{1},k_{2};\tau^{\prime}\right)e^{-\left(\beta_{\text{D}k_{1}}+\beta_{\text{D}k_{2}}\right)\left(2t_{1}-\tau-\tau^{\prime}\right)},
\end{align}
\begin{align}
 \mathcal{P}_{\text{I}}=&\int\text{d}k_{1}\text{d}k_{2}\int_{t_{0}}^{t_{1}}\text{d}\tau\text{d}\tau^{\prime}\phi\left(k_{1},k_{2};\tau\right)\nonumber\\
&\quad\times\phi^{\ast}\left(k_{1},k_{2};\tau^{\prime}\right)e^{-\beta_{\text{D}k_{1}}\left(2t_{1}-\tau-\tau^{\prime}\right)}e^{-\beta_{\text{D}k_{2}}\left|\tau-\tau^{\prime}\right|} \nonumber \\
&+\int\text{d}k_{1}\text{d}k_{2}\int_{t_{0}}^{t_{1}}\text{d}\tau\text{d}\tau^{\prime}\phi\left(k_{1},k_{2};\tau\right)\nonumber\\
&\quad\times\phi^{\ast}\left(k_{1},k_{2};\tau^{\prime}\right)e^{-\beta_{\text{D}k_{1}}\left|\tau-\tau^{\prime}\right|}e^{-\beta_{\text{D}k_{2}}\left(2t_{1}-\tau-\tau^{\prime}\right)}-2\mathcal{P}_{\text{II}},
\end{align}
 and
\begin{align}
\mathcal{P}_{\text{vac}}=&\int\text{d}k_{1}\text{d}k_{2}\int_{t_{0}}^{t_{1}}\text{d}\tau\text{d}\tau^{\prime}\phi\left(k_{1},k_{2};\tau\right)\nonumber\\
&\times\phi^{\ast}\left(k_{1},k_{2};\tau^{\prime}\right)e^{-\left(\beta_{\text{D}k_{1}}+\beta_{\text{D}k_{2}}\right)\left|\tau-\tau^{\prime}\right|}-\mathcal{P}_{\text{I}}-\mathcal{P}_{\text{II}},
\end{align}
 or
\begin{align}
&\int\text{d}k_{1}\text{d}k_{2}\int_{t_{0}}^{t_{1}}\text{d}\tau\text{d}\tau^{\prime}\phi\left(k_{1},k_{2};\tau\right)\nonumber\\
&\quad\times\phi^{\ast}\left(k_{1},k_{2};\tau^{\prime}\right)e^{-\left(\beta_{\text{D}k_{1}}+\beta_{\text{D}k_{2}}\right)\left|\tau-\tau^{\prime}\right|}=1. \label{eq:norm}
\end{align}
We interpret the normalization condition \eqref{eq:norm} as a constraint on the temporal distance over which biphoton wave functions can interfere due to interaction with the reservoir.  
 
\subsection{$\omega$-space expressions}

The density operator derived at the end of the previous Section \eqref{eq:rhogenk}
can also be understood in the frequency representation. To get there,
as in \cite{Yang:2008}, we take
\begin{align}
a_{m\omega}\equiv&\sqrt{\frac{\text{d}k_{m}\left(\omega\right)}{\text{d}\omega}}a_{mk_{m}\left(\omega\right)}=\left[v_{m}\left(\omega\right)\right]^{-1/2}a_{mk_{m}\left(\omega\right)},\\
\phi_{\text{P}}\left(\omega\right)\equiv&\sqrt{\frac{\text{d}k_{\text{P}}\left(\omega\right)}{\text{d}\omega}}\phi_{\text{P}}\left(k_{\text{P}}\left(\omega\right)\right)=\left[v_{\text{P}}\left(\omega\right)\right]^{-1/2}\phi_{\text{P}}\left(k_{\text{P}}\left(\omega\right)\right),
\end{align}
and
\begin{align}
\phi\left(\omega_{1},\omega_{2};\tau\right)\equiv & \sqrt{\frac{\text{d}k_{\text{D}}\left(\omega_{2}\right)}{\text{d}\omega}}\sqrt{\frac{\text{d}k_{\text{D}}\left(\omega_{2}\right)}{\text{d}\omega}}\phi\left(k_{\text{D}}\left(\omega_{1}\right),k_{\text{D}}\left(\omega_{2}\right);\tau\right)\nonumber \\
= & \left[v_{\text{D}}\left(\omega_{1}\right)v_{\text{D}}\left(\omega_{2}\right)\right]^{-1/2}\phi\left(k_{\text{D}}\left(\omega_{1}\right),k_{\text{D}}\left(\omega_{2}\right);\tau\right),
\end{align}
where the derivatives have been introduced to ensure normalization
in frequency space. Furthermore, we approximate the effective area
$\mathcal{A}$ and group velocities $v_{m}$ as being constant over
the frequency ranges of interest \cite{Yang:2008}, and take $t_0=-t_1$. While earlier work
took $t_{0}\rightarrow-\infty$, $t_{1}\rightarrow\infty$ \cite{Yang:2008},
doing so here would result in scattering loss occurring for all times.
Thus, as mentioned above, as a first approximation we take $t_{0}=-L/\left(2v_{\text{P}}\right)$,
$t_{1}=L/\left(2v_{\text{P}}\right)$ where $L$ is the length of
the waveguide, and $v_{\text{P}}$ is the group velocity of the pump
field. As shown in Appendix \ref{sec:classical} \eqref{eq:alpha}, the usual attenuation
coefficients $\alpha_{m}\left(\omega\right)$ are related to our $\beta_{m}\left(\omega\right)$
via $\alpha_{m}\left(\omega\right)=2\beta_{m}\left(\omega\right)/v_{m}$.
This equality combined with our choice of interaction duration leads
to ratios of group velocities, $r\equiv v_\text{D}/v_\text{P}$, appearing in expressions written in
terms of $\alpha_{m}\left(\omega\right)$. Recalling \eqref{eq:S},
the two-photon term \eqref{eq:pk2} can be written as 
\begin{align}
\rho_{2}=&\frac{1}{2}\int\text{d}\omega_{1}\text{d}\omega_{2}\text{d}\omega_{1}^{\prime}\text{d}\omega_{2}^{\prime}\theta_{2}\left(\omega_{1},\omega_{2},\omega_{1}^{\prime},\omega_{2}^{\prime}\right)e^{-\left[\alpha_{\text{D}}\left(\omega_{1}\right)+\alpha_{\text{D}}\left(\omega_{2}\right)\right]Lr/4}\nonumber\\
&\times e^{-\left[\alpha_{\text{D}}\left(\omega_{1}^{\prime}\right)+\alpha_{\text{D}}\left(\omega_{2}^{\prime}\right)\right]Lr/4}a_{\text{D}\omega_{1}}^{\dagger}a_{\text{D}\omega_{2}}^{\dagger}\left\vert \text{vac}\right\rangle \left\langle \text{vac}\right\vert a_{\text{D}\omega_{1}^{\prime}}a_{\text{D}\omega_{2}^{\prime}},\label{rhoII}
\end{align}
the one-photon term as
\begin{align}
\rho_{1}=&2\int\text{d}\omega_{1}\text{d}\omega_{2}\text{d}\omega\left[\theta_{1}\left(\omega_{1},\omega,\omega_{2},\omega\right)e^{-\left[\alpha_{\text{D}}\left(\omega_{1}\right)+\alpha_{\text{D}}\left(\omega_{2}\right)\right]Lr/4}\right.\nonumber\\
&-\left.\theta_{2}\left(\omega_{1},\omega,\omega_{2},\omega\right)e^{-\left[\alpha_{\text{D}}\left(\omega_{1}\right)+\alpha_{\text{D}}\left(\omega_{2}\right)+2\alpha_{\text{D}}\left(\omega\right)\right]Lr/4}\right]\nonumber\\
&\quad\times a_{\text{D}\omega_{1}}^{\dagger}\left\vert \text{vac}\right\rangle \left\langle \text{vac}\right\vert a_{\text{D}\omega_{2}},\label{rhoI}
\end{align}
and the vacuum term as
\begin{align}
\rho_{0}=&\int\text{d}\omega_{1}\text{d}\omega_{2}\left[\theta_{0}\left(\omega_{1},\omega_{2},\omega_{1},\omega_{2}\right)\right.\nonumber\\
&-\theta_{1}\left(\omega_{1},\omega_{2},\omega_{1},\omega_{2}\right)\left(e^{-\alpha_{\text{D}}\left(\omega_{1}\right)Lr/2}+e^{-\alpha_{\text{D}}\left(\omega_{2}\right)Lr/2}\right)\nonumber\\
&+\left.\theta_{2}\left(\omega_{1},\omega_{2},\omega_{1},\omega_{2}\right)e^{-\left[\alpha_{\text{D}}\left(\omega_{1}\right)+\alpha_{\text{D}}\left(\omega_{2}\right)\right]Lr/2}\right]\left\vert \text{vac}\right\rangle \left\langle \text{vac}\right\vert.\label{rhovac}
\end{align}

The three different density operator wave functions differ only in terms of their temporal integrals:
\begin{widetext}
\begin{align}
\theta_{2}\left(\omega_{1},\omega_{2},\omega_{3},\omega_{4}\right)=&A\left(\omega_{1},\omega_{2}\right)A^{*}\left(\omega_{3},\omega_{4}\right)\int\text{d}\omega\text{d}\omega^{\prime}B\left(\omega_{1},\omega_{2},\omega\right)B^{*}\left(\omega_{3},\omega_{4},\omega^{\prime}\right)\nonumber\\
&\times\int_{-L/\left(2v_{\text{P}}\right)}^{L/\left(2v_{\text{P}}\right)}\frac{\text{d}\tau\text{d}\tau^{\prime}}{4\pi^{2}}e^{\left(i\omega_{1}+i\omega_{2}-i\omega+\alpha_{\text{D}}\left(\omega_{1}\right)v_{\text{D}}/2+\alpha_{\text{D}}\left(\omega_{2}\right)v_{\text{D}}/2-\alpha_{\text{P}}\left(\omega\right)v_{\text{P}}/2\right)\tau}\nonumber\\
&\quad\times e^{-\left(i\omega_{3}+i\omega_{4}-i\omega^{\prime}-\alpha_{\text{D}}\left(\omega_{3}\right)v_{\text{D}}/2-\alpha_{\text{D}}\left(\omega_{4}\right)v_{\text{D}}/2+\alpha_{\text{P}}\left(\omega^{\prime}\right)v_{\text{P}}/2\right)\tau^{\prime}},\label{eq:BWF}
\end{align}
\begin{align}
\theta_{1}\left(\omega_{1},\omega_{2},\omega_{3},\omega_{4}\right)=&A\left(\omega_{1},\omega_{2}\right)A^{*}\left(\omega_{3},\omega_{4}\right)\int\text{d}\omega\text{d}\omega^{\prime}B\left(\omega_{1},\omega_{2},\omega\right)B^{*}\left(\omega_{3},\omega_{4},\omega^{\prime}\right)\nonumber\\
&\times\int_{-L/\left(2v_{\text{P}}\right)}^{L/\left(2v_{\text{P}}\right)}\frac{\text{d}\tau\text{d}\tau^{\prime}}{4\pi^{2}}e^{\left(i\omega_{1}+i\omega_{2}-i\omega+\alpha_{\text{D}}\left(\omega_{1}\right)v_{\text{D}}/2-\alpha_{\text{P}}\left(\omega\right)v_{\text{P}}/2\right)\tau}\nonumber\\
&\quad\times e^{-\left(i\omega_{3}+i\omega_{4}-i\omega^{\prime}-\alpha_{\text{D}}\left(\omega_{3}\right)v_{\text{D}}/2+\alpha_{\text{P}}\left(\omega^{\prime}\right)v_{\text{P}}/2\right)\tau^{\prime}}e^{-\left[\alpha_{\text{D}}\left(\omega_{2}\right)+\alpha_{\text{D}}\left(\omega_{4}\right)\right]v_{\text{D}}\left|\tau-\tau^{\prime}\right|/4},
\end{align}
\begin{align}
\theta_{0}\left(\omega_{1},\omega_{2},\omega_{3},\omega_{4}\right)=&A\left(\omega_{1},\omega_{2}\right)A^{*}\left(\omega_{3},\omega_{4}\right)\int\text{d}\omega\text{d}\omega^{\prime}B\left(\omega_{1},\omega_{2},\omega\right)B^{*}\left(\omega_{3},\omega_{4},\omega^{\prime}\right)\nonumber\\
&\times\int_{-L/\left(2v_{\text{P}}\right)}^{L/\left(2v_{\text{P}}\right)}\frac{\text{d}\tau\text{d}\tau^{\prime}}{4\pi^{2}}e^{\left(i\omega_{1}+i\omega_{2}-i\omega-\alpha_{\text{P}}\left(\omega\right)v_{\text{P}}/2\right)\tau}e^{-\left(i\omega_{3}+i\omega_{4}-i\omega^{\prime}+\alpha_{\text{P}}\left(\omega^{\prime}\right)v_{\text{P}}/2\right)\tau^{\prime}}\nonumber\\
&\quad\times e^{-\left[\alpha_{\text{D}}\left(\omega_{1}\right)+\alpha_{\text{D}}\left(\omega_{2}\right)+\alpha_{\text{D}}\left(\omega_{3}\right)+\alpha_{\text{D}}\left(\omega_{4}\right)\right]v_{\text{D}}\left|\tau-\tau^{\prime}\right|/4},
\end{align}
\end{widetext}
where
\begin{equation}
A\left(\omega_{1},\omega_{2}\right)=\frac{\mu}{\nu}\frac{i\overline{\chi}_{2}L}{2\overline{n}^{3}v_{\text{D}}\sqrt{v_{\text{P}}\mathcal{A}}}\sqrt{\frac{\hbar\omega_{1}\omega_{2}}{2\pi\varepsilon_{0}}},
\end{equation}
and
\begin{align}
B\left(\omega_{1},\omega_{2},\omega\right)=&\sqrt{\omega}\phi_{\text{P}}\left(\omega\right)e^{-\alpha_{\text{P}}\left(\omega\right)L/4}\nonumber\\
&\times\frac{\sin\left[\left(k_{\text{D}}\left(\omega_{1}\right)+k_{\text{D}}\left(\omega_{2}\right)-k_{\text{P}}\left(\omega\right)\right)L/2\right]}{\left(k_{\text{D}}\left(\omega_{1}\right)+k_{\text{D}}\left(\omega_{2}\right)-k_{\text{P}}\left(\omega\right)\right)L/2}.\label{eq:B}
\end{align}
In fact, we note that if there are no scattering losses in the D mode, $\theta_{2}\left(\omega_{1},\omega_{2},\omega_{3},\omega_{4}\right)=\theta_{1}\left(\omega_{1},\omega_{2},\omega_{3},\omega_{4}\right)=\theta_{0}\left(\omega_{1},\omega_{2},\omega_{3},\omega_{4}\right)$, and $\rho_{1}=\rho_{0}=0$.

\section{Photon wave functions and exit probabilities}

\label{sec:BWF}

\subsection{Exit probabilities}

In the lossless case, in the limit of a low probability of pair production, the state that is generated with probability $\left\vert \zeta\right\vert ^{2}$ per pump pulse contains two photons that exit the nonlinear region of the waveguide without scattering with unit probability. Working in the same regime in our current calculation, however, for a $\rho_\text{gen}$ that is generated with probability $\left\vert \zeta\right\vert ^{2}$, there is only a probability
\begin{equation}
\mathcal{P}_{2}=\int\text{d}\omega_{1}\text{d}\omega_{2}\theta_{2}\left(\omega_{1},\omega_{2},\omega_{1},\omega_{2}\right)e^{-\left[\alpha_{\text{D}}\left(\omega_{1}\right)+\alpha_{\text{D}}\left(\omega_{2}\right)\right]Lr/2},\label{eq:II}
\end{equation}
that both photons of the pair exit the waveguide without scattering. There is also a
probability 
\begin{align}
\mathcal{P}_{1}= & 2\int\text{d}\omega_{1}\text{d}\omega_{2}\left(\theta_{1}\left(\omega_{1},\omega_{2},\omega_{1},\omega_{2}\right)e^{-\alpha_{\text{D}}\left(\omega_{1}\right)Lr/2}\right.\nonumber \\
 & -\left.\theta_{2}\left(\omega_{1},\omega_{2},\omega_{1},\omega_{2}\right)e^{-\left[\alpha_{\text{D}}\left(\omega_{1}\right)+\alpha_{\text{D}}\left(\omega_{2}\right)\right]Lr/2}\right),\label{eq:I}
\end{align}
that only one photon of the pair exits, and a probability 
\begin{align}
\mathcal{P}_{0}= & \int\text{d}\omega_{1}\text{d}\omega_{2}\left(\theta_{0}\left(\omega_{1},\omega_{2},\omega_{1},\omega_{2}\right)-2\theta_{1}\left(\omega_{1},\omega_{2},\omega_{1},\omega_{2}\right)e^{-\alpha_{\text{D}}\left(\omega_{1}\right)Lr/2}\right.\nonumber \\
 & +\left.\theta_{2}\left(\omega_{1},\omega_{2},\omega_{1},\omega_{2}\right)e^{-\left[\alpha_{\text{D}}\left(\omega_{1}\right)+\alpha_{\text{D}}\left(\omega_{2}\right)\right]Lr/2}\right),
\end{align}
that neither exits. For a fixed $\left\vert\zeta\right\vert^2$, $r\approx 1$, and fixed density operator wave functions, $\theta_i$, the physics of these probabilities is clear:  1) With an increasing product of device length and downconverted mode loss the probability of two photons exiting the device in the waveguide mode \eqref{eq:II} always decreases. 2) As there is an additional $\alpha_{\text{D}}\left(\omega\right)$ in the decaying exponential coefficient in front of the negative $\theta_{2}\left(\omega_{1},\omega_{2},\omega_{1},\omega_{2}\right)$
in \eqref{eq:I} compared to the decaying exponential coefficient in front of the positive $\theta_{1}\left(\omega_{1},\omega_{2},\omega_{1},\omega_{2}\right)$,
the probability of a single photon exiting the device in the guided mode first increases and then decreases as $\alpha_\text{D}L$ increases from zero. For the same reason, the ratio of the probability of single photons to pairs exiting the waveguide, $\mathcal{P}_2/\mathcal{P}_1$, also decreases as $\alpha_\text{D}L$ increases. 3) Finally, as $\alpha_\text{D}L$ tends to infinity $\mathcal{P}_{0}$ tends to 1 (see \cite{Dezfouli:2014} for related results in a $\chi_3$ structure).  Along with our normalization condition \eqref{eq:norm} expressed in $\omega$-space
\begin{equation}
\int\text{d}\omega_1\text{d}\omega_2\theta_{0}\left(\omega_{1},\omega_{2},\omega_{1},\omega_{2}\right)=1,
\end{equation}
the form of $\mathcal{P}_0$ ensures that the three probabilities sum to one regardless of device length or downconverted mode loss, as the trace of a density operator must.  To
first order in the nonlinearity, which we identify with the undepleted pump approximation, and in the limit of a low probability of pair
production, these results are exact and enable many calculations, which we now explore.

\subsection{General expressions}

A comparison between the biphoton wave function associated with the two-photon term and the usual biphoton wave function that results in the absence of scattering loss is most easily seen in the frequency representation. Rewriting \eqref{rhoII} in a form that makes its pure state nature explicit
\begin{equation}
\rho_2 = C_2^\dagger\left\vert \text{vac}\right\rangle \left\langle \text{vac}\right\vert C_2,
\end{equation}
where
\begin{align}
C_2^\dagger =& \frac{1}{\sqrt{2}}\int\text{d}\omega_{1}\text{d}\omega_{2}\,A\left(\omega_{1},\omega_{2}\right)e^{-\left[\alpha_{\text{D}}\left(\omega_{1}\right)+\alpha_{\text{D}}\left(\omega_{2}\right)\right]Lr/4}\nonumber\\
&\times\int\text{d}\omega B\left(\omega_{1},\omega_{2},\omega\right)\int_{-L/\left(2v_{\text{P}}\right)}^{L/\left(2v_{\text{P}}\right)}\frac{\text{d}\tau}{2\pi}\nonumber\\
&\quad\times e^{\left[i\omega_{1}+i\omega_{2}-i\omega+\alpha_{\text{D}}\left(\omega_{1}\right)v_{\text{D}}/2+\alpha_{\text{D}}\left(\omega_{2}\right)v_{\text{D}}/2-\alpha_{\text{P}}\left(\omega\right)v_{\text{P}}/2\right]\tau}a_{\text{D}\omega_{1}}^{\dagger}a_{\text{D}\omega_{2}}^{\dagger},
\end{align}
performing the temporal integral, and recalling \eqref{eq:B}, enables identification of
\begin{widetext}
\begin{align}
\Phi_{\text{2}}\left(\omega_{1},\omega_{2}\right)\equiv & A\left(\omega_{1},\omega_{2}\right)\int\text{d}\omega\,\sqrt{\omega}\phi_{\text{P}}\left(\omega\right)\frac{\sin\left[\left(k_{\text{D}}\left(\omega_{1}\right)+k_{\text{D}}\left(\omega_{2}\right)-k_{\text{P}}\left(\omega\right)\right)L/2\right]}{\left(k_{\text{D}}\left(\omega_{1}\right)+k_{\text{D}}\left(\omega_{2}\right)-k_{\text{P}}\left(\omega\right)\right)L/2}\nonumber \\
 & \times\frac{\sinh\left\{ \left[i\omega_{1}+i\omega_{2}-i\omega+\alpha_{\text{D}}\left(\omega_{1}\right)v_{\text{D}}/2+\alpha_{\text{D}}\left(\omega_{2}\right)v_{\text{D}}/2-\alpha_{\text{P}}\left(\omega\right)v_{\text{P}}/2\right]L/\left(2v_{\text{P}}\right)\right\} }{\pi\left[i\omega_{1}+i\omega_{2}-i\omega+\alpha_{\text{D}}\left(\omega_{1}\right)v_{\text{D}}/2+\alpha_{\text{D}}\left(\omega_{2}\right)v_{\text{D}}/2-\alpha_{\text{P}}\left(\omega\right)v_{\text{P}}/2\right]}\nonumber \\
 & \times e^{-\alpha_{\text{P}}\left(\omega\right)L/4}e^{-\left[\alpha_{\text{D}}\left(\omega_{1}\right)+\alpha_{\text{D}}\left(\omega_{2}\right)\right]Lr/4},\label{eq:lossyBWF}
\end{align}
\end{widetext}
as the biphoton wave function associated with the two-photon contribution to the density matrix describing SPDC in a lossy nonlinear waveguide. Comparing this biphoton wave function
with the biphoton wave function that results in a lossless nonlinear waveguide \eqref{eq:PhiNoLoss}, 
\begin{align}
\phi\left(\omega_{1},\omega_{2}\right)= & A\left(\omega_{1},\omega_{2}\right)\int\text{d}\omega\,\sqrt{\omega}\phi_{\text{P}}\left(\omega\right)\nonumber \\
&\times\frac{\sin\left[\left(k_{\text{D}}\left(\omega_{1}\right)+k_{\text{D}}\left(\omega_{2}\right)-k_{\text{D}}\left(\omega\right)\right)L/2\right]}{\left(k_{\text{D}}\left(\omega_{1}\right)+k_{\text{D}}\left(\omega_{2}\right)-k_{\text{P}}\left(\omega\right)\right)L/2}\nonumber \\
 & \times\frac{\sinh\left[\left(i\omega_{1}+i\omega_{2}-i\omega\right)L/\left(2v_{\text{P}}\right)\right]}{\pi\left(i\omega_{1}+i\omega_{2}-i\omega\right)},
\end{align}
two new features can be noticed. The first, seen on the final line
of \eqref{eq:lossyBWF}, is the appearance of exponential decay terms associated with
the scattering loss of pump and generated photons as they traverse
the waveguide. The second, and perhaps less expected, is that the
hyperbolic sine term now contains loss coefficients.  In the absence of loss and the limit of extending the interaction time to infinity, $L/\left(2v_\text{P}\right)\rightarrow\infty$, we would have
\begin{equation}
\frac{\sinh\left[\left(i\omega_{1}+i\omega_{2}-i\omega\right)L/\left(2v_{\text{P}}\right)\right]}{\pi\left(i\omega_{1}+i\omega_{2}-i\omega\right)}\approx\delta\left(\omega_{1}+\omega_{2}-\omega\right),\label{eq:Dirac}
\end{equation}
expressing energy conservation. Once loss is added, this conservation condition, which essentially still holds in the absence of loss even for a finite interaction time if it is long enough, is modified.  The loss coefficients in the new version of \eqref{eq:Dirac} appearing in \eqref{eq:lossyBWF} in the presence of loss lead to the ``loss matching'' term mentioned in the introduction.  Approximating the frequency integral of \eqref{eq:lossyBWF} as (cf. \eqref{eq:Dirac}) 
\begin{widetext}
\begin{align}
&\Phi\left(\omega_1,\omega_2\right)\nonumber\\
&\approx A\left(\omega_1,\omega_2\right) \sqrt{\omega_{1}+\omega_{2}} \phi_{\text{P}}\left(\omega_{1}+\omega_{2}\right)e^{-\alpha_{\text{P}}\left(\omega_{1}+\omega_{2}\right)L/4}e^{-\left[\alpha_{\text{D}}\left(\omega_{1}\right)+\alpha_{\text{D}}\left(\omega_{2}\right)\right]Lr/4}\nonumber \\
 & \times\frac{\sin\left( \left\{k_{\text{D}}\left(\omega_{1}\right)+k_{\text{D}}\left(\omega_{2}\right)-k_{\text{P}}\left(\omega_{1}+\omega_{2}\right)-i\left[ \alpha_{\text{D}}\left(\omega_{1}\right)r+\alpha_{\text{D}}\left(\omega_{2}\right)r-\alpha_{\text{P}}\left(\omega_1+\omega_2\right)\right]/2\right\}L/2\right) }{\left\{k_{\text{D}}\left(\omega_{1}\right)+k_{\text{D}}\left(\omega_{2}\right)-k_{\text{P}}\left(\omega_{1}+\omega_{2}\right)-i\left[ \alpha_{\text{D}}\left(\omega_{1}\right)r+\alpha_{\text{D}}\left(\omega_{2}\right)r-\alpha_{\text{P}}\left(\omega_1+\omega_2\right)\right]/2\right\}L/2},
\end{align}
we find
\begin{align}
&\left|\Phi_{\text{2}}\left(\omega_{1},\omega_{2}\right)\right|^{2} \nonumber \\
&=\left|A\left(\omega_{1},\omega_{2}\right)\right|^{2}\left|\phi_{\text{P}}\left(\omega_{1}+\omega_{2}\right)\right|^{2}\left(\omega_{1}+\omega_{2}\right)e^{-\alpha_{\text{P}}\left(\omega_{1}+\omega_{2}\right)L/2}e^{-\left[\alpha_{\text{D}}\left(\omega_{1}\right)+\alpha_{\text{D}}\left(\omega_{2}\right)\right]Lr/2} \nonumber \\
&\times\frac{\sin^{2}\left\{ \left[k_{\text{D}}\left(\omega_{1}\right)+k_{\text{D}}\left(\omega_{2}\right)-k_{\text{P}}\left(\omega_{1}+\omega_{2}\right)\right]L/2\right\} +\sinh^{2}\left\{ \left[\alpha_{\text{D}}\left(\omega_{1}\right)r+\alpha_{\text{D}}\left(\omega_{2}\right)r-\alpha_{\text{P}}\left(\omega_1+\omega_2\right)\right]L/4\right\} }{\left\{ \left[k_{\text{D}}\left(\omega_{1}\right)+k_{\text{D}}\left(\omega_{2}\right)-k_{\text{P}}\left(\omega_{1}+\omega_{2}\right)\right]L/2\right\} ^{2}+\left\{ \left[\alpha_{\text{D}}\left(\omega_{1}\right)r+\alpha_{\text{D}}\left(\omega_{2}\right)r-\alpha_{\text{P}}\left(\omega_1+\omega_2\right)\right]L/4\right\} ^{2}}.
\label{eq:ToPlot}
\end{align}
\end{widetext}
We note that this expression bears a strong resemblance to that for the power generated in a classical sum frequency generation (SFG) process with undepleted pumps that includes scattering loss in the continuous wave (CW) limit (see \cite{Sutherland:2003} and Appendix \ref{sec:classical}).

\subsection{Biphoton probability densities}

Here we plot the biphoton probability density $\left\vert \Phi_{2}\left(\omega_{1},\omega_{2}\right)\right\vert ^{2}$
associated with coincidence counts \eqref{eq:ToPlot} for three qualitatively distinct
cases. As an example, we consider the effects
of including loss in the calculation of previously predicted biphoton
probability densities in Bragg reflection waveguides \cite{Zhukovsky:2012}.
We Taylor expand 
\begin{equation}
k_{m}\left(\omega\right)\approx k_{m_{0}}+\left(\omega-\omega_{m}\right)/v_{m}+\Lambda_{m}\left(\omega-\omega_{m}\right)^{2},
\end{equation}
where $k_{\text{P}_{0}}=k_{\text{P}}\left(\omega_{\text{P}}\right)$,
$k_{\text{D}_{0}}=k_{\text{D}}\left(\omega_{\text{P}}/2\right)$,
$v_{\text{P}}=\text{d}k_{\text{P}}\left(\omega\right)/\left.\text{d}\omega\right|_{\omega=\omega_{\text{P}}}$,
$v_{\text{D}}=\text{d}k_{\text{D}}\left(\omega\right)/\left.\text{d}\omega\right|_{\omega=\omega_\text{P}/2}$,
$2\Lambda_{\text{P}}=\text{d}^{2}k_{\text{P}}\left(\omega\right)/\left.\text{d}\omega^{2}\right|_{\omega=\omega_{\text{P}}}$,
$2\Lambda_{\text{D}}=\text{d}^{2}k_{\text{D}}\left(\omega\right)/\left.\text{d}\omega^{2}\right|_{\omega=\omega_\text{P}/2}$,
and $k_{\text{P}_{0}}=2k_{\text{D}_{0}}$. The relevant parameters are $v_{\text{P}}=74.3$
$\mu$m/ps, $v_{\text{D}}=89.8$ $\mu$m/ps, $\Lambda_{\text{P}}=2.92\times10^{-6}$
ps$^{2}$/$\mu$m, $\Lambda_{\text{D}}=7.07\times10^{-7}$ ps$^{2}$/$\mu$m. Furthermore, we take
$L=2$ mm, and a Gaussian pump pulse waveform
\begin{equation}
\phi_{\text{P}}\left(\omega\right)=\frac{\exp\left(\frac{-\left(\omega-\omega_{\text{P}}\right)^{2}}{2\Delta^{2}}\right)}{\left(\sqrt{\pi}\Delta\right)^{1/2}},
\end{equation}
with an intensity full width at half maximum (FWHM) in time of $T=2\sqrt{\ln\left(2\right)}/\Delta=20$
fs, and $\omega_\text{P}=2\pi c /\left( 775\,\text{nm}\right)$. If losses associated with each mode are frequency independent and $\alpha_{\text{P}}v_{\text{P}}\approx2\alpha_{\text{D}}v_{\text{D}}$
the shape of the biphoton probability density is exactly as in the
absence of loss: the exponential attenuation factor alters the
number of generated photons, but not the photon pair frequency correlations
(see Fig. 1a) 
\begin{align}
&\left|\Phi_{\text{2}}^{\text{Balanced}}\left(\omega_{1},\omega_{2}\right)\right|^{2}\propto\left|\phi_{\text{P}}\left(\omega_{1}+\omega_{2}\right)\right|^{2}\nonumber \\
&\quad\times\frac{\sin\left[\left(k_{\text{D}}\left(\omega_{1}\right)+k_{\text{D}}\left(\omega_{2}\right)-k_{\text{P}}\left(\omega_{1}+\omega_{2}\right)\right)L/2\right]^{2}}{\left[\left(k_{\text{D}}\left(\omega_{1}\right)+k_{\text{D}}\left(\omega_{2}\right)-k_{\text{P}}\left(\omega_{1}+\omega_{2}\right)\right)L/2\right]^{2}}.
\end{align}
If instead frequency independent losses are such that $\alpha_{\text{P}}v_{\text{P}}\gg2\alpha_{\text{D}}v_{\text{D}}$
the biphoton probability density is near Lorentzian 
\begin{align}
&\left|\Phi_{\text{2}}^{\text{Unbalanced}}\left(\omega_{1},\omega_{2}\right)\right|^{2}\propto\left|\phi_{\text{P}}\left(\omega_{1}+\omega_{2}\right)\right|^{2}\nonumber \\
&\quad\times\frac{1}{\left[\left(k_{\text{D}}\left(\omega_{1}\right)+k_{\text{D}}\left(\omega_{2}\right)-k_{\text{P}}\left(\omega_{1}+\omega_{2}\right)\right)L/2\right]^{2}+\left[\alpha_{\text{P}}L/4\right]^{2}},
\end{align}
a shape that we plot in Fig. 1b for $\alpha_{\text{P}}=40$ cm$^{-1}$, $\alpha_{\text{D}}=2 \text{ cm}^{-1}$ \cite{Abolhasem:2009}. (The large difference in the size of these loss coefficients for that experiment is due to the different guiding mechanisms employed at the pump (second harmonic) frequency versus the downconverted (fundamental) frequency.  In this Bragg reflection waveguide, the D mode is guided by total internal reflection, while the P mode is guided by Bragg reflections.) Using \eqref{eq:II} and \eqref{eq:I}, we see that compared to the lossless case, when all photon pairs generated per pump pulse $\left|\zeta\right|^{2}$ exit the waveguide, here only $0.3\left|\zeta\right|^{2}$ do as pairs, while $0.5\left|\zeta\right|^{2}$ do as singles. Although the temporal integrals can be performed analytically, the integrals over frequency cannot, and so were calculated on a 64$\times$64$\times$64$\times$64 grid with $\omega_1$ and $\omega_2$ ranging from $\omega_\text{P}/2-5\Delta_1$ to $\omega_\text{P}/2+5\Delta_1$ and $\omega$ and $\omega^\prime$ ranging from $\omega_\text{P}-5\Delta_1$ to $\omega_\text{P}+5\Delta_1$ with $\Delta_1=2\sqrt{\ln{2}}/\left(20\text{ fs}\right)$. Additionally, the associated Schmidt number \cite{Law:2004}, $K$, characterizing the frequency correlations of the biphoton wave functions, has been reduced from 618 to 180 as we have moved moved from balanced losses to large SH losses and the side lobes of the phase matching sinc function contribution to the biphoton probability density have become washed out.  The Schmidt number quantifies the effective number of frequency modes that contribute to $\Phi_2\left(\omega_1,\omega_2\right)$. We note that this behaviour is similar to that achieved intentionally in earlier works~\cite{Branczyk:2011,BenDixon:2013}, only here it is simply a consequence of waveguide loss and comes with an associated reduction in photon pair waveguide exit probability as well as an increase in the ratio of single photon to pair exit probability. Lastly, we consider the mathematically interesting possibility of losses further reducing the frequency correlations of a biphoton probability density having a reasonably small associated $K$ to begin with. For this we consider the same structure as above, with a pump pulse duration of $T=2$ ps and a quadratic frequency dependent loss model.  When losses are balanced the biphoton probability density looks as in Fig. 1c., whereas for the quadratic loss profile,
$\alpha_{\text{D}}\left(\omega\right)=1.77\times 10^{10}\left(1/2-\omega/\omega_\text{P}\right)^2$ cm$^{-1}$, with $\alpha_{\text{P}}\left(\omega\right)=0$, it takes the shape shown in Fig. 1d.  For this loss profile, the generation bandwidth has been greatly reduced, much as when strong filters are applied to achieve nearly frequency uncorrelated photons.  The interesting loss feature has led to a biphoton wave function that is naturally nearly frequency uncorrelated, with an associated Schmidt number of only 1.29 compared to the balanced loss case of 76.3.  Again we use a 64$\times$64$\times$64$\times$64 grid, here with $\omega_1$ and $\omega_2$  ranging from $\omega_\text{P}/2-5\Delta_1$ to $\omega_\text{P}/2+5\Delta_1$ and $\omega$ and $\omega^\prime$ ranging from $\omega_\text{P}-5\Delta_2$ to $\omega_\text{P}+5\Delta_2$ with $\Delta_2=2\sqrt{\ln{2}}/\left(2\text{ ps}\right)$, to calcualte exit probabilities. However here only $0.004\left|\zeta\right|^{2}$ exit the waveguide as pairs, while $0.01\left|\zeta\right|^{2}$ do as singles, leading to a very small $\mathcal{P}_2/\mathcal{P}_1$.

\begin{figure}[hbt]
\includegraphics[trim= 15mm 0mm 0mm 0mm, clip, width=0.6\textwidth]{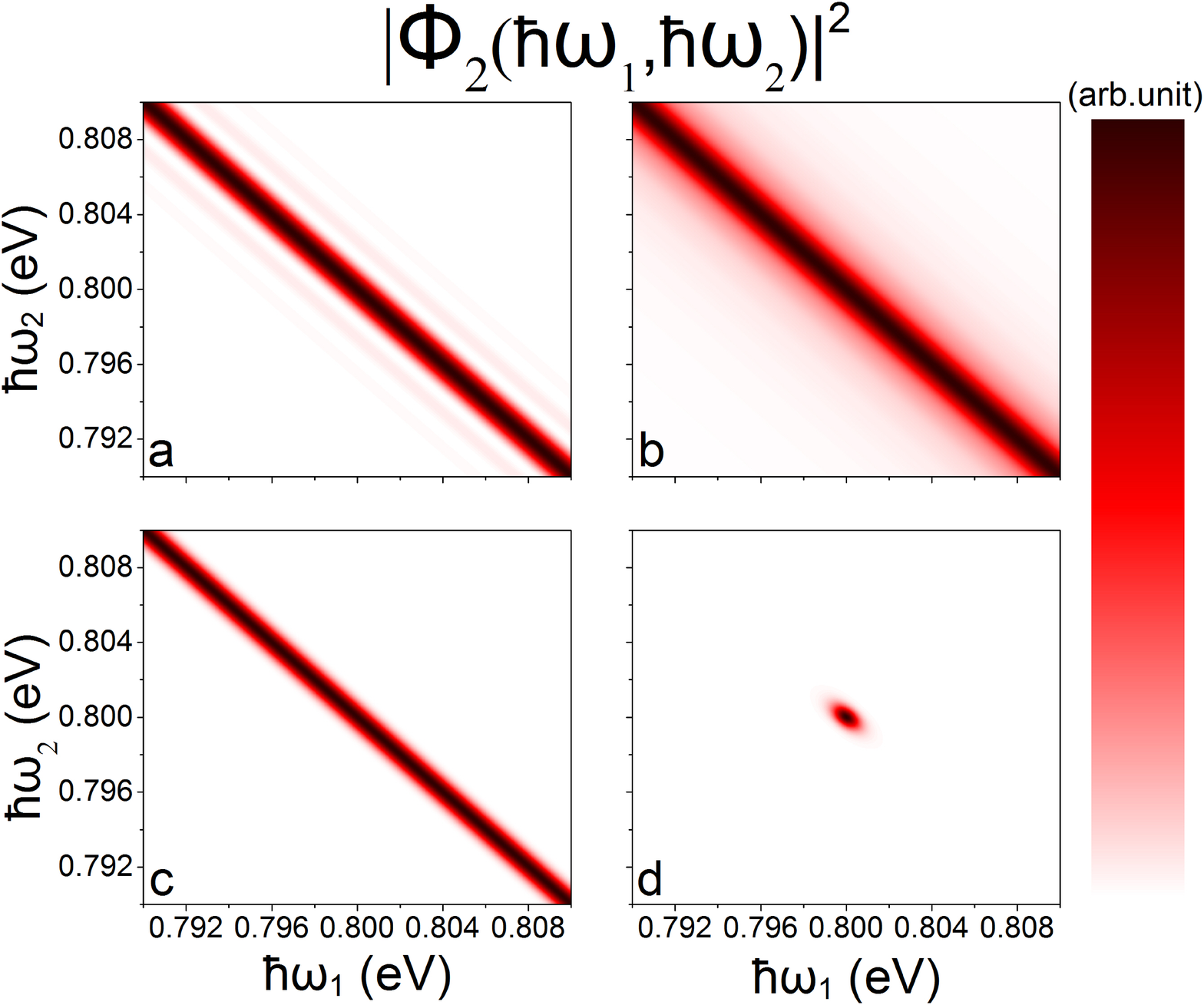}
\caption{Lossy biphoton probability densities. Figures a. and b. show balanced loss and large loss at the second harmonic, respectively, for $T=20$ fs. Figures c. and d. show balanced loss and the quadratic loss profile $\alpha_{\text{D}}\left(\omega\right)=1.77\times 10^{10}\left(1/2-\omega/\omega_\text{P}\right)^2 \text{ cm}^{-1}$, respectively, for $T=2$ ps.}
\end{figure}

\section{Conclusions}

\label{sec:conclusions}

In conclusion, we have presented a formalism capable of handling SPDC and linear scattering loss concurrently in a waveguide. A density operator describing the state of generated photons was calculated and shown to be composed of a two photon, a single photon, and a vacuum piece, with traces that sum to 1 corresponding to the probabilities of two photons, one photon, or zero photons exiting the device.  In general, the biphoton wave function associated with the two photon contribution in a lossy waveguide was shown to additionally exhibit both exponential decay terms and a loss matching hyperbolic sine term compared to the usual lossless biphoton wave function. By looking at biphoton probability densities in three different regimes we demonstrated that it is possible to imagine losses such that they do not affect the shape of the biphoton probability density whatsoever, that losses can wash out the side lobes of the phase matching sinc function contribution to the biphoton probability density, and also that losses might possibly be controlled to engineer desired biphoton probability densities, at the cost of reducing the probability of a pair exiting the waveguide as well as increasing the ratio of the probability of single photons to pairs exiting the waveguide. 

\begin{acknowledgments}
This work was supported in part by the ARC Centre for Ultrahigh bandwidth
Devices for Optical Systems (CUDOS) (project number CE110001018),
the Natural Sciences and Engineering Research Council of Canada, and
the Macquarie University Research Fellowship Scheme.
\end{acknowledgments}
\appendix

\section{Relevant commutation relations}
\label{sec:comms}
Here we evaluate the two commutations relations necessary to perform the trace leading to \eqref{eq:trace}.  In particular
\begin{align}
&\left[a_{mk}e^{-\beta_{mk}\left(t_{1}-\tau\right)},a_{m^{\prime}k^{\prime}}^{\dagger}e^{-\beta_{m^{\prime}k^{\prime}}\left(t_{1}-\tau^{\prime}\right)}\right]\nonumber\\
&=\delta_{mm^{\prime}}\delta\left(k-k^{\prime}\right)e^{-\beta_{mk}\left(2t_{1}-\tau-\tau^{\prime}\right)},
\end{align}
and
\begin{align}
&\left[\int_{t}^{t_{1}}\text{d}\tau F_{mk}\left(\tau\right)e^{-\beta_{mk}\left(\tau-t\right)},\int_{t^{\prime}}^{t_{1}}\text{d}\tau^{\prime}F_{m^{\prime}k^{\prime}}^{\dagger}\left(\tau^{\prime}\right)e^{-\beta_{m^{\prime}k^{\prime}}\left(\tau^{\prime}-t^{\prime}\right)}\right]\nonumber\\
&=2\beta_{mk}\delta_{mm^{\prime}}\delta\left(k-k^{\prime}\right)\nonumber\\
&\quad\times\int_{t}^{t_{1}}\text{d}\tau\int_{t^{\prime}}^{t_{1}}\text{d}\tau^{\prime}\delta\left(t-t^{\prime}\right)e^{-\beta_{mk}\left(\tau+\tau^{\prime}-t-t^{\prime}\right)}\nonumber\\
&=\frac{\beta_{mk}\delta_{mm^{\prime}}\delta\left(k-k^{\prime}\right)}{\pi}\nonumber\\
&\quad\times\int_{t}^{t_{1}}\text{d}\tau\int_{t^{\prime}}^{t_{1}}\text{d}\tau^{\prime}\int_{-\infty}^{\infty}\text{d}\omega e^{-i\omega\left(\tau-\tau^{\prime}\right)}e^{-\beta_{mk}\left(\tau+\tau^{\prime}-t-t^{\prime}\right)}\nonumber\\
&=\frac{\beta_{mk}\delta_{mm^{\prime}}\delta\left(k-k^{\prime}\right)e^{\beta_{mk}\left(t+t^{\prime}\right)}}{\pi}\nonumber\\
&\quad\times\int_{-\infty}^{\infty}\text{d}\omega\int_{t}^{t_{1}}\text{d}\tau e^{-\tau\left(\beta_{mk}+i\omega\right)}\int_{t^{\prime}}^{t_{1}}\text{d}\tau^{\prime}e^{-\tau^{\prime}\left(\beta_{mk}-i\omega\right)}\nonumber\\
&=\frac{\beta_{mk}\delta_{mm^{\prime}}\delta\left(k-k^{\prime}\right)e^{\beta_{mk}\left(t+t^{\prime}\right)}}{\pi}\nonumber\\
&\quad\times\int_{-\infty}^{\infty}\text{d}\omega\frac{e^{-t\left(\beta_{mk}+i\omega\right)}-e^{-t_{1}\left(\beta_{mk}+i\omega\right)}}{\beta_{mk}+i\omega}\nonumber\\
&\quad\quad\times\frac{e^{-t^{\prime}\left(\beta_{mk}-i\omega\right)}-e^{-t_{1}\left(\beta_{mk}-i\omega\right)}}{\beta_{mk}-i\omega}\nonumber\\
&=\delta_{mm^{\prime}}\delta\left(k-k^{\prime}\right)e^{\beta_{mk}\left(t+t^{\prime}\right)}\nonumber\\
&\quad\times\left[e^{-\beta_{mk}\left(t+t^{\prime}\right)}e^{-\beta_{mk}\left|t-t^{\prime}\right|}-e^{-\beta_{mk}\left(t+t_{1}\right)}e^{-\beta_{mk}\left(t_{1}-t\right)}\right.\nonumber\\
&\quad\quad\left.-e^{-\beta_{mk}\left(t^{\prime}+t_{1}\right)}e^{-\beta_{mk}\left(t_{1}-t^{\prime}\right)}+e^{-2\beta_{mk}t_{1}}\right]\nonumber\\
&=\delta_{mm^{\prime}}\delta\left(k-k^{\prime}\right)\left(e^{-\beta_{mk}\left|t-t^{\prime}\right|}-e^{-\beta_{mk}\left(2t_{1}-t-t^{\prime}\right)}\right),
\end{align}
where the second to last equality follows from Eq. 3.354-5 of \cite{Gradshteyn:2007}.

\section{Classical calculation}
\label{sec:classical}

Following the formalism presented in an earlier work~\cite{Helt:2012},
we begin with the general Hamiltonian 
\begin{equation}
H=H_{\text{L}}+H_{\text{NL}}+H_{\text{R}}+H_{\text{C}},\label{eq:HFull}
\end{equation}
where the linear, nonlinear, reservoir, and coupling pieces are defined
above in \eqref{eq:HL}, \eqref{eq:H_NL}, \eqref{eq:HR}, and \eqref{eq:HC},
respectively. With the CW limit in mind, we then Taylor expand dispersion
relations in the linear Hamiltonian $\omega_{mk}\approx\omega_{m}+v_{m}\left(k-k_{m}\right)$,
and approximate $\omega\approx\omega_{mk}$ $\Omega_{m\mu k}\approx\Omega_{m\mu}$,
$c_{m\mu k}\approx c_{m\mu}$, and $d_{mk}^{i}\left(x,y\right)\approx d_{m}^{i}\left(x,y\right)$ in
the others. Finally, introducing effective field operators
\begin{equation}
g_{m}\left(z,t\right)=\int\frac{\text{d}k}{\sqrt{2\pi}}a_{mk}e^{i\left(k-k_{m}\right)z},
\end{equation}
\begin{equation}
h_{m\mu}\left(z,t\right)=\int\frac{\text{d}k}{\sqrt{2\pi}}b_{m\mu k}e^{i\left(k-k_{m}\right)z},
\end{equation}
and noting that, for CW fields, the points at which the $k$'s appearing
in the $g_{m}$'s are most naturally expanded about are well-separated,
we introduce two new $m$ labels ($m=\text{S}$ for signal, $m=\text{I}$
for idler) corresponding to what was formerly just the fundamental (downconverted) mode $m=\text{D}$,
with
\begin{equation}
\left[g_{\text{S}},g_{\text{I}}^{\dagger}\right]=0,
\end{equation}
and relabel what was called the pump mode in our SPDC calculations as $m=\text{SH}$ for second harmonic.
This allows us to rewrite our initial Hamiltonian \eqref{eq:HFull}
as 
\begin{align}
H_{\text{L}}=&\sum_{m=\text{S,I,SH}}\left\{ \hbar\omega_{m}\int\text{d}zg_{m}^{\dagger}g_{m}\right.\nonumber \\
&+\left.\frac{i}{2}\hbar v_{m}\int\text{d}z\left(\frac{\partial g_{m}^{\dagger}}{\partial z}g_{m}-g_{m}^{\dagger}\frac{\partial g_{m}}{\partial z}\right)\right\} , \\
H_{\text{NL}}=&-\gamma\int\text{d}ze^{i\left(k_{\text{S}}+k_{\text{I}}-k_{\text{SH}}\right)z}g_{\text{SH}}^{\dagger}g_{\text{S}}g_{\text{I}}+\text{H.c.}, \\
H_{\text{R}}=&\sum_{m=\text{S,I},\text{SH}}\hbar\int\text{d}z\text{d}\mu\Omega_{m\mu}h_{m\mu}^{\dagger}h_{m\mu}, \\
H_{\text{I}}=&\sum_{m=\text{S,I},\text{SH}}\hbar\int\text{d}z\text{d}\mu\left(c_{m\mu}g_{m}^{\dagger}h_{m\mu}+c_{m\mu}^{\ast}h_{m\mu}^{\dagger}g_{m}\right),
\end{align}
where
\begin{equation}
\gamma=2\sqrt{\frac{\left(\hbar\omega_{\text{S}}\right)\left(\hbar\omega_{\text{I}}\right)\left(\hbar\omega_{\text{SH}}\right)}{2^{3}\varepsilon_{0}\mathcal{A}}}\frac{\overline{\chi}_{2}}{\overline{n}^{3}},
\end{equation}
and
\begin{equation}
\mathcal{A}=\left|\int\text{d}x\text{d}y\frac{\bar{n}^{3}\chi_{2}^{ijk}d_{\text{S}}^{i}\left(x,y\right)d_{\text{I}}^{j}\left(x,y\right)\left[d_{\text{SH}}^{k}\left(x,y\right)\right]^{*}}{\bar{\chi}_{2}\varepsilon_{0}^{3/2}n^{2}\left(x,y;\omega_{\text{S}}\right)n^{2}\left(x,y;\omega_{\text{I}}\right)n^{2}\left(x,y;\omega_{\text{SH}}\right)}\right|^{-2}.
\end{equation}
The Heisenberg equations of motion yield
\begin{align}
\frac{\partial h_{m\mu}}{\partial t}=&  -\frac{i}{\hbar}\left[h_{m\mu},H\right]\nonumber \\
=&  -i\Omega_{m\mu}h_{m\mu}-ic_{m\mu}^{*}g_{m}, \\
\frac{\partial g_{\text{SH}}}{\partial t} =& -\frac{i}{\hbar}\left[g_{\text{SH}},H\right]\nonumber \\
 =& -i\omega_{\text{SH}}g_{\text{SH}}-v_{\text{SH}}\frac{\partial g_{\text{SH}}}{\partial z}+\frac{i\gamma}{\hbar}e^{i\left(k_{\text{S}}+k_{\text{I}}-k_{\text{SH}}\right)z}g_{\text{S}}g_{\text{I}}\nonumber \\
&-i\int\text{d}\mu c_{\text{SH}\mu}h_{\text{SH}\mu}, \\
\frac{\partial g_{\text{S}}}{\partial t} =& -\frac{i}{\hbar}\left[g_{\text{S}},H\right]\nonumber \\
=& -i\omega_{\text{S}}g_{\text{S}}-v_{\text{S}}\frac{\partial g_{\text{S}}}{\partial z}+\frac{i\gamma^{*}}{\hbar}e^{-i\left(k_{\text{S}}+k_{\text{I}}-k_{\text{SH}}\right)z}g_{\text{I}}^{\dagger}g_{\text{SH}}\nonumber \\
&-i\int\text{d}\mu c_{\text{S}\mu}h_{\text{S}\mu}, \\
\frac{\partial g_{\text{I}}}{\partial t} =& -\frac{i}{\hbar}\left[g_{\text{I}},H\right]\nonumber \\
=& -i\omega_{\text{I}}g_{\text{I}}-v_{\text{I}}\frac{\partial g_{\text{I}}}{\partial z}+\frac{i\gamma^{*}}{\hbar}e^{-i\left(k_{\text{S}}+k_{\text{I}}-k_{\text{SH}}\right)z}g_{\text{S}}^{\dagger}g_{\text{SH}}\nonumber \\
&-i\int\text{d}\mu c_{\text{I}\mu}h_{\text{I}\mu},
\end{align}
the first of which has solution 
\begin{equation}
h_{m\mu}\left(z,t\right)=h_{m\mu}\left(z,0\right)e^{-i\Omega_{m\mu}t}-i\int_{0}^{t}g_{m}\left(z,\tau\right)c_{m\mu}^{\ast}e^{-i\Omega_{m\mu}\left(t-\tau\right)}\text{d}\tau,\label{eq:h}
\end{equation}
assuming that the interaction is switched on at $t=0$ (i.e. light
enters the waveguide at $t=0$). Following arguments presented in
the main text, we approximate approximate $\left\vert c_{m\mu}\right\vert ^{2}\approx \mathcal{C}_{m}$
as independent of $\mu$ and write $\text{d}\mu=\left(\text{d}\mu/\text{d}\Omega_{m\mu}\right)\text{d}\Omega_{m\mu}$,
also approximating the density of states $\text{d}\mu/\text{d}\Omega_{m\mu}\approx\mathcal{D}_{m}$,
and substitute \eqref{eq:h} into the equations for the $g_{m}$,
yielding 
\begin{align}
\frac{\partial g_{\text{SH}}}{\partial t}=&-i\omega_{\text{SH}}g_{\text{SH}}-v_{\text{SH}}\frac{\partial g_{\text{SH}}}{\partial z}+\frac{i\gamma}{\hbar}e^{i\left(k_{\text{S}}+k_{\text{I}}-k_{\text{SH}}\right)z}g_{\text{S}}g_{\text{I}}\nonumber \\
&-\beta_{\text{SH}}g_{\text{SH}}-r_{\text{SH}}, \\
\frac{\partial g_{\text{S}}}{\partial t}=&-i\omega_{\text{S}}g_{\text{S}}-v_{\text{S}}\frac{\partial g_{\text{S}}}{\partial z}+\frac{i\gamma^{*}}{\hbar}e^{-i\left(k_{\text{S}}+k_{\text{I}}-k_{\text{SH}}\right)z}g_{\text{I}}^{\dagger}g_{\text{SH}}\nonumber \\
&-\beta_{\text{S}}g_{\text{S}}-r_{\text{S}},\\
\frac{\partial g_{\text{I}}}{\partial t}=&-i\omega_{\text{I}}g_{\text{I}}-v_{\text{I}}\frac{\partial g_{\text{I}}}{\partial z}+\frac{i\gamma^{*}}{\hbar}e^{-i\left(k_{\text{S}}+k_{\text{I}}-k_{\text{SH}}\right)z}g_{\text{S}}^{\dagger}g_{\text{SH}}\nonumber \\
&-\beta_{\text{I}}g_{\text{I}}-r_{\text{I}},
\end{align}
where we have defined
\begin{equation}
r_{m}\left(z,t\right)=i\int\text{d}\mu h_{m\mu}\left(z,0\right)c_{m\mu}e^{-i\Omega_{m\mu}t},
\end{equation}
and the fixed $k$ version of our loss rate above as $\beta_{m}=\mathcal{C}_{m}\pi\mathcal{D}_{m}.$
We note that the commutation relation for this real-space fluctuation
operator is exactly what might be expected from its $k$-space analogue
\eqref{eq:FComm} 
\begin{equation}
\left[r_{m}\left(z,t\right),r_{m^{\prime}}^{\dagger}\left(z^{\prime},t^{\prime}\right)\right]=2\beta_{m}\delta_{mm^{\prime}}\delta\left(t-t^{\prime}\right)\delta\left(z-z^{\prime}\right)
\end{equation}
We then put 
\begin{align}
g_{m}&=\widetilde{g}_{m}e^{-i\omega_{m}t}, \\
\omega_{\text{SH}}&=\omega_{\text{S}}+\omega_{\text{I}},
\end{align}
and write the equations above in terms of new operators
\begin{align}
G_{m}&=\sqrt{\hbar\omega_{m}v_{m}}\widetilde{g}_{m}, \\
R_{m}&=\sqrt{\frac{\hbar\omega_{m}}{v_{m}}}e^{i\omega_{m}t}r_{m},
\end{align}
such that $G_{m}^{\dagger}G_{m}=P_{m}$ has units of power:
\begin{widetext}
\begin{align}
\frac{1}{v_{\text{SH}}}\frac{\partial G_{\text{SH}}}{\partial t}+\frac{\partial G_{\text{SH}}}{\partial z}=&\frac{i\gamma}{\hbar v_{\text{SH}}}\sqrt{\frac{\omega_{\text{SH}}v_{\text{SH}}}{\hbar\omega_{\text{S}}v_{\text{S}}\omega_{\text{I}}v_{\text{I}}}}e^{i\left(k_{\text{S}}+k_{\text{I}}-k_{\text{SH}}\right)z}G_{\text{S}}G_{\text{I}}\-\frac{\beta_{\text{SH}}}{v_{\text{SH}}}G_{\text{SH}}-R_{\text{SH}}, \\
\frac{1}{v_{\text{S}}}\frac{\partial G_{\text{S}}}{\partial t}+\frac{\partial G_{\text{S}}}{\partial z}=&\frac{i\gamma^{*}}{\hbar v_{\text{S}}}\sqrt{\frac{\omega_{\text{S}}v_{\text{S}}}{\hbar\omega_{\text{I}}v_{\text{I}}\omega_{\text{SH}}v_{\text{SH}}}}e^{-i\left(k_{\text{S}}+k_{\text{I}}-k_{\text{SH}}\right)z}G_{\text{I}}^{\dagger}G_{\text{SH}}-\frac{\beta_{\text{S}}}{v_{\text{S}}}G_{\text{S}}-R_{\text{S}}, \\
\frac{1}{v_{\text{I}}}\frac{\partial G_{\text{I}}}{\partial t}+\frac{\partial G_{\text{I}}}{\partial z}=&\frac{i\gamma^{*}}{\hbar v_{\text{I}}}\sqrt{\frac{\omega_{\text{I}}v_{\text{I}}}{\hbar\omega_{\text{S}}v_{\text{S}}\omega_{\text{SH}}v_{\text{SH}}}}e^{-i\left(k_{\text{S}}+k_{\text{I}}-k_{\text{SH}}\right)z}G_{\text{S}}^{\dagger}G_{\text{SH}}-\frac{\beta_{\text{I}}}{v_{\text{I}}}G_{\text{I}}-R_{\text{I}},
\end{align}
\end{widetext}
We work in the undepleted pump approximation, in the limit of stationary
fields, where the time derivatives vanish, and the limit of strong
pumps $G_{\text{S,I}}\gg G_{\text{SH}}$, leaving
\begin{align}
\frac{\partial G_{\text{SH}}}{\partial z}=&\frac{i\gamma}{\hbar v_{\text{SH}}}\sqrt{\frac{\omega_{\text{SH}}v_{\text{SH}}}{\hbar\omega_{\text{S}}v_{\text{S}}\omega_{\text{I}}v_{\text{I}}}}e^{i\left(k_{\text{S}}+k_{\text{I}}-k_{\text{SH}}\right)z}G_{\text{S}}G_{\text{I}}\nonumber \\
&-\frac{\beta_{\text{SH}}}{v_{\text{SH}}}G_{\text{SH}}-R_{\text{SH}}, \\
\frac{\partial G_{\text{S}}}{\partial z}=&-\frac{\beta_{\text{S}}}{v_{\text{S}}}G_{\text{S}}-R_{\text{S}}, \\
\frac{\partial G_{\text{I}}}{\partial z}=&-\frac{\beta_{\text{I}}}{v_{\text{I}}}G_{\text{I}}-R_{\text{I}}.
\end{align}
Tracing over reservoir operators at zero temperature, we find that
$\left\langle R_{m}\right\rangle _{R}=0$. Writing $\left\langle G_{m}\right\rangle _{R}=G_{m}$
for notational simplicity, and recognizing 
\begin{equation}
\alpha_{m}=2\frac{\beta_{m}}{v_{m}},\label{eq:alpha}
\end{equation}
we arrive at coupled mode equations
\begin{align}
\frac{\partial G_{\text{SH}}}{\partial z}&=\frac{i\gamma}{\hbar v_{\text{SH}}}\sqrt{\frac{\omega_{\text{SH}}v_{\text{SH}}}{\hbar\omega_{\text{S}}v_{\text{S}}\omega_{\text{I}}v_{\text{I}}}}e^{i\left(k_{\text{S}}+k_{\text{I}}-k_{\text{SH}}\right)z}G_{\text{S}}G_{\text{I}}-\frac{\alpha_{\text{SH}}}{2}G_{\text{SH}}, \\
\frac{\partial G_{\text{S}}}{\partial z}&=-\frac{\alpha_{\text{S}}}{2}G_{\text{S}}, \\
\frac{\partial G_{\text{I}}}{\partial z}&=-\frac{\alpha_{\text{I}}}{2}G_{\text{I}}.
\end{align}
For a waveguide extending from $z=-L/2$ to
$z=L/2$ we find
\begin{align}
&G_{\text{SH}}\left(L/2\right)=i\frac{G_{\text{S}}\left(-L/2\right)G_{\text{I}}\left(-L/2\right)}{\mathcal{\sqrt{PA}}}\nonumber\\
&\times\frac{e^{\left(ik_{\text{S}}+ik_{\text{I}}-ik_{\text{SH}}-\alpha_{\text{S}}-\alpha_{\text{I}}\right)L/2}-e^{-\left(ik_{\text{S}}+ik_{\text{I}}-ik_{\text{SH}}+\alpha_{\text{SH}}\right)L/2}}{i\left(k_{\text{S}}+k_{\text{I}}-k_{\text{SH}}\right)-\frac{\alpha_{\text{I}}+\alpha_{\text{S}}-\alpha_{\text{SH}}}{2}},
\end{align}
where $\mathcal{P}=2\varepsilon_0 \bar{n}^6v_\text{S}v_\text{I}v_\text{SH}/\left[\left(\bar{\chi}_2\right)^2\omega_\text{SH}^2\right]$. Setting $P_{\text{SH}}=G_{\text{SH}}^{\dagger}\left(L/2\right)G_{\text{SH}}\left(L/2\right),\\$
$P_{\text{S}}=G_{\text{S}}^{\dagger}\left(-L/2\right)G_{\text{S}}\left(-L/2\right)$, and
$P_{\text{I}}=G_{\text{I}}^{\dagger}\left(-L/2\right)G_{\text{I}}\left(-L/2\right)$, we arrive at the well-known result \cite{Sutherland:2003}
\begin{align}
&P_{\text{SH}}=P_{\text{S}}P_{\text{I}}\frac{L^{2}}{\mathcal{PA}}e^{-\left(\alpha_{\text{SH}}+\alpha_{\text{S}}+\alpha_{\text{I}}\right)L/2}\nonumber \\
&\times\frac{\sin^{2}\left[\left(k_{\text{S}}+k_{\text{I}}-k_{\text{SH}}\right)L/2\right]+\sinh^{2}\left[\left(\alpha_{\text{S}}+\alpha_{\text{I}}-\alpha_{\text{SH}}\right)L/4\right]}{\left[\left(k_{\text{S}}+k_{\text{I}}-k_{\text{SH}}\right)L/2\right]^{2}+\left[\left(\alpha_{\text{S}}+\alpha_{\text{I}}-\alpha_{\text{SH}}\right)L/4\right]^{2}}.
\end{align}
Note how the exponential, sine, and hyperbolic sine terms are essentially the same here and in \eqref{eq:ToPlot}.

\bibliography{loss}

\end{document}